\documentclass[aps,prd,notitlepage,showpacs,nofootinbib,superscriptaddress]{revtex4-1}

\usepackage{graphicx}
\usepackage[utf8]{inputenc}

\usepackage{dsfont}
\usepackage{amsmath}
\usepackage{amssymb}
\usepackage{amsfonts}
\usepackage{float}
\usepackage{comment}
\usepackage{slashed}
\usepackage{bbm}
\usepackage{bbold}

\usepackage{booktabs}

\usepackage{mathtools,braket,blkarray}

\usepackage{amsmath}
\usepackage{amssymb}

\usepackage[usenames,dvipsnames]{color}
\usepackage[colorinlistoftodos]{todonotes}
\usepackage[colorlinks=true,citecolor=darkred,urlcolor=darkred, pdfborder={0 0 0}]{hyperref}

\usepackage{multirow}
\definecolor{darkred}{rgb}{0.6,0,0}
\definecolor{darkgreen}{rgb}{0.0, 0.4, 0.0}
\usepackage[colorinlistoftodos]{todonotes}

\definecolor{linkcolor}{rgb}{0,0,0.5}

\newcommand {\ignore}[1]{}

\def \znbb {$\rm 0\nu\beta\beta$ }

\def\gsim{\raise0.3ex\hbox{$\;>$\kern-0.75em\raise-1.1ex\hbox{$\sim\;$}}}
\def\lsim{\raise0.3ex\hbox{$\;<$\kern-0.75em\raise-1.1ex\hbox{$\sim\;$}}}
\def\lfv{lepton flavour violation }
\def\lnv{lepton number violating }

\newcommand{\sm}{{Standard Model }}

\usepackage{soul}

\definecolor{mightnightblue}{RGB}{25,25,112}

\definecolor{brown}{rgb}{0.59, 0.29, 0.0}

\def\lfv{lepton flavour violation }
 
\def\lnv{lepton number violation }

\def\21{$\mathrm{SU(2)_L \otimes U(1)_Y}$}
\def\lfv{lepton flavor violation }

\def\lnv{lepton number violation }
\def\sm{standard model }

\newcommand{\AddrAHEP}{%
  AHEP Group, Institut de F\'{i}sica Corpuscular --
  C.S.I.C./Universitat de Val\`{e}ncia, Parc Cient\'ific de Paterna.\\
 C/ Catedr\'atico Jos\'e Beltr\'an, 2 E-46980 Paterna (Valencia) - SPAIN}

\begin{document}

\begin{flushright}
  {\tt USTC-ICTS/PCFT-20-06}
\end{flushright}

\title{\boldmath \color{BrickRed} Predictions from warped flavordynamics based on the $T'$ family group}

\author{Peng Chen}\email{pche@mail.ustc.edu.cn}
\affiliation{College of Information Science and Engineering,Ocean University of China, Qingdao 266100, China}
\author{Gui-Jun Ding}\email{dinggj@ustc.edu.cn}
\affiliation{
Peng Huanwu Center for Fundamental Theory, Hefei, Anhui 230026, China}
\affiliation{
Interdisciplinary Center for Theoretical Study and Department of Modern Physics, \\
University of Science and Technology of China, Hefei, Anhui 230026, China}
\author{Jun-Nan Lu}\email{JunNan.Lu@ific.uv.es}
\affiliation{\AddrAHEP}
\author{Jos\'{e} W. F. Valle}\email{valle@ific.uv.es}
\affiliation{\AddrAHEP}

\begin{abstract}
\vspace{0.5cm}

We propose a realistic theory of fermion masses and mixings using a five-dimensional warped scenario where all fermions propagate in the bulk and the Higgs field is localized on the IR brane.
The assumed $T'$ flavor symmetry is broken on the branes by flavon fields, providing a consistent scenario where fermion mass hierarchies arise from adequate choices of the bulk mass parameters,
while quark and lepton mixing angles are restricted by the family symmetry.
Neutrino mass splittings, mixing parameters and the Dirac CP phase all arise from the type-I seesaw mechanism and are tightly correlated,
  leading to predictions for the neutrino oscillation parameters,
  as well as expected \znbb decay rates within reach of upcoming experiments.
The scheme also provides a good global description of flavor observables in the quark sector.

\end{abstract}

\maketitle
\noindent

\section{Introduction}
\label{Sect:intro}

Understanding flavor from first principles is one of the greatest challenges in particle physics. The coin has two sides. On the one hand there is the problem of understanding the observed hierarchies of quark and lepton masses,
explaining why is the muon about 200 times heavier than the electron, or why does the top quark seem to play such a special role in being the heaviest.

On the other hand comes the problem of finding a \textit{rationale} for the observed pattern of mixing parameters.
This problem has only become trickier after the discovery of neutrino oscillations~\cite{McDonald:2016ixn,Kajita:2016cak} which implies not only the need for neutrino masses -- and understanding their smallness with respect to the charged fermion masses --
but also the need to understand why the pattern of neutrino mixing is so special when compared to that of quarks~\cite{deSalas:2017kay}.

The Standard Model (SM) lacks an organizing principle to account for the observed flavor properties.
The existence of flat extra dimensions has been suggested as a way to shed light on the possible nature of the family symmetry~\cite{Altarelli:2005yp}.
In particular, six-dimensional theories compactified on a torus have been suggested~\cite{deAnda:2018oik,deAnda:2018yfp} and a successful
model has recently been proposed~\cite{deAnda:2019jxw,deAnda:2020pti} in which fermions are nicely arranged within the framework of an $A_4$ family symmetry, with good predictions for fermion masses and mixings,
including the ``golden'' quark-lepton unification formula~\cite{Morisi:2011pt,King:2013hj,Morisi:2013eca,Bonilla:2014xla,Reig:2018ocz}.
Although intriguingly successful, this orbifold theory of flavor~\cite{deAnda:2019jxw,deAnda:2020pti} remains far from giving a complete description of mass hierarchies.

As a possible alternative to the flat-extra-dimensions approach here we turn to the possibility of warped extra dimensions.
These have been proposed by Randall \& Sundrum~\cite{Randall:1999ee} in order to address the hierarchy problem without the need to invoke supersymmetry.
The fundamental scale of gravity gets exponentially reduced with respect to the Planck scale by having the Higgs sector localized near the boundary of the extra dimensions.
Here we assume the \sm fermions to propagate in the bulk, though peaked towards either brane.
This allows us to address at once both aspects of the flavor problem: the fermion mass hierarchy problem, as well as their mixing pattern, with the help of a family symmetry group. This follows the general approach suggested in Ref.~\cite{Chen:2015jta}.
In such scenarios fermion mass hierarchies are accounted for by adequate choices of the bulk mass parameters, while quark and lepton mixing angles are restricted by the assumed family symmetry, broken on the branes by flavon fields.

Our present scenario employs a $T'$-based family group and predicts the neutrino mixing parameters and the Dirac CP violation phase in terms of only two independent parameters at leading order.
$T'$ is the double covering of $A_4$.
  Besides the triplet representation $\mathbf{3}$ and the three singlet representations $\mathbf{1}$, $\mathbf{1'}$ and $\mathbf{1''}$, in common with $A_4$, it has three doublet representations
  $\mathbf{2}$, $\mathbf{2'}$ and $\mathbf{2''}$.
  We will exploit the presence of the doublet representations to describe the quark sector, by assigning the three quark families to a reducible singlet plus doublet representation.
%
%
In contrast to Ref.~\cite{Chen:2015jta} where neutrinos were Dirac particles, here a viable description of neutrino oscillations requires neutrinos to be Majorana particles. Moreover, given the predicted regions for the oscillation parameters, it follows that there must be a lower bound on the neutrinoless double beta decay rate even if the spectrum is normal-ordered.
We show that the model also provides a successful global description of flavor, consistent with the observed CKM quark mixing matrix, in which the successful Gatto-Sartori relation emerges in leading order.


The paper is organised as follows.
After sketching the theoretical framework in Sec.~\ref{sec:preliminaries} we move on to describe the lepton sector in Sec.~\ref{Sec:lepton},
and the quark sector in Sec.~\ref{sec:quarks}, giving the corresponding field content and quantum numbers.
In Sec.~\ref{sec:analysis} we give a numerical analysis of the resulting flavor predictions.
The sub-leading corrections to the mass terms and mixing parameters are studied in Sec.~\ref{sec:highorder}.
  Finally, in Sec.~\ref{sec:bulked_higgs} we comment briefly on a variant construction in which the Higgs lives in the bulk.
  In Sec.~\ref{sec:Conclusions} we conclude, giving complementary material in the Appendices, as follows.
  The group theory of $T'$ is summarized in Appendix~\ref{sec:Tp_group}.
  The 5-D profiles of fields are presented in Appendix~\ref{sec:XD},
  while in Appendix~\ref{sec:vacuum_alignment}, we investigate the vacuum alignment of the flavon fields.

\section{theoretical Preliminaries}
\label{sec:preliminaries}

\begin{figure}[h!]
\centering
\begin{tabular}{cc}
\includegraphics[width=0.5\linewidth]{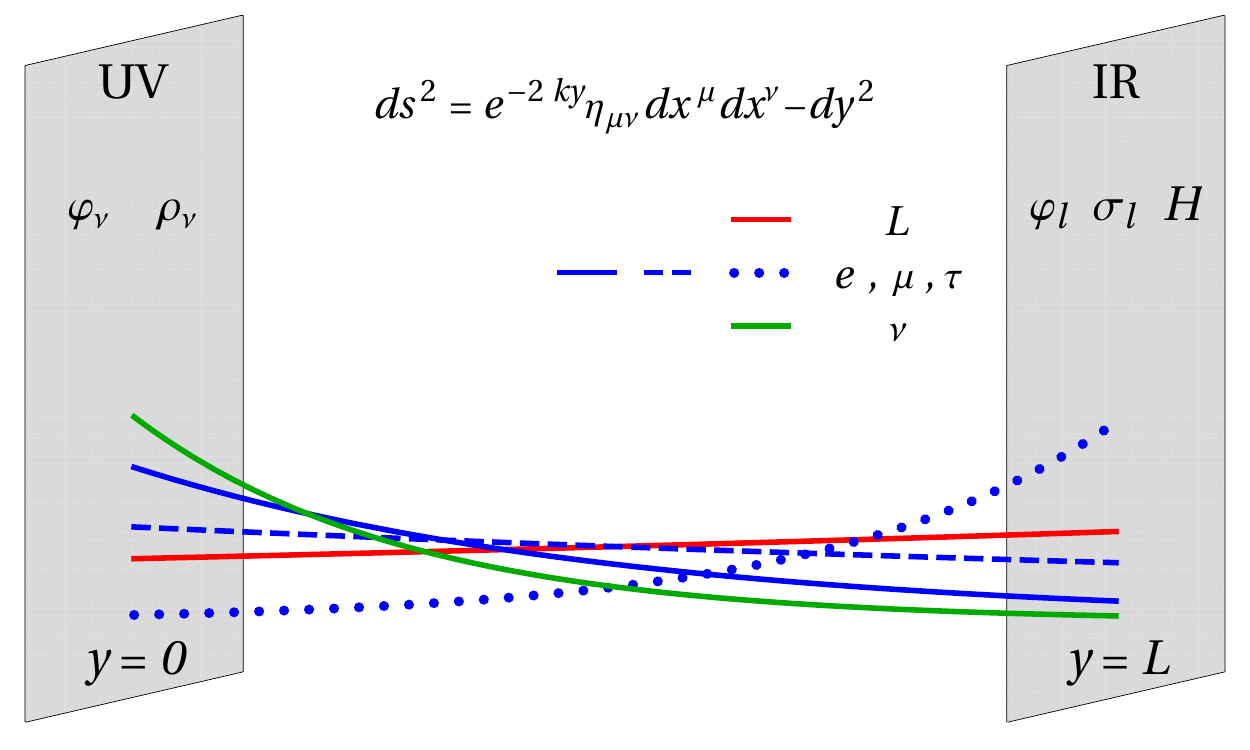}
\end{tabular}
\caption{Zero-mode profiles of the lepton fields in the fifth dimension. The flavon fields $\varphi_l$, $\sigma_l$ and $\varphi_{\nu}$, $\rho_\nu$ are localized on the IR and UV branes, respectively. }
\label{fig:wavef_lepton}
\end{figure}

Here we study the implementation of a flavor symmetry within a warped extra dimensional theory context.
The bulk electroweak gauge symmetry is extended to $SU(2)_L\otimes SU(2)_{R}\otimes U(1)_{B-L}$, ensuring consistency with restrictions from the electroweak precision measurements~\cite{Agashe:2003zs}.
We denote the gauge fields and gauge couplings associated with the gauge groups $SU(2)_L$, $SU(2)_R$ and $U(1)_{B-L}$ as $W^a_{L\mu}$, $W^a_{R\mu}$, $X_{\mu}$ and $g_{L}$, $g_{R}$, $g_X$ respectively, with $a=1, 2, 3$.
The extended electroweak gauge group $SU(2)_L\otimes SU(2)_{R}\otimes U(1)_{B-L}$ is broken down to the Standard Model (SM) group $SU(2)_{L}\times U(1)_Y$ by orbifold boundary condition on the UV brane~\cite{Agashe:2003zs}.
This symmetry breaking pattern can be achieved by the following assignment of boundary conditions\footnote{These boundary conditions can be naturally obtained by adding a $SU(2)_R$ scalar doublet or triplet field on the UV brane~\cite{Csaki:2003dt,Csaki:2005vy}, if they acquire non-zero vacuum expectation value (VEV). },
\begin{equation}
W_{L\mu}^{1,2,3}(++),~~~
B_{\mu}(++),~~~W_{R\mu}^{1,2}(-+),~~~Z'_{\mu}(-+)\,,
\end{equation}
where the first (second) sign in the bracket stands for the boundary condition on the UV (IR) brane, and ``+'' (``$-$'') refers to the
Neumann (Dirichlet) boundary condition. The fields $B_{\mu}$ and $Z'_{\mu}$ are linear combinations of the original fields $W^{3}_{R\mu}$ and $X_{\mu}$
\begin{equation}
\begin{pmatrix}
B_{\mu} \\
Z'_{\mu}
\end{pmatrix}=\frac{1}{\sqrt{g^2_R+g^2_X}}\begin{pmatrix}
 g_{R}    ~&  g_X \\
-g_X    ~&   g_R
\end{pmatrix}\begin{pmatrix}
X_{\mu} \\
W^{3}_{R\mu}
\end{pmatrix}\,.
\end{equation}
The hypercharge coupling of $U(1)_Y$ is given by
\begin{equation}
g_Y=\frac{g_Rg_X}{\sqrt{g^2_R+g^2_X}}\,.
\end{equation}
Only the fields with $(++)$ boundary condition have zero modes upon the Kaluza-Klein (KK) decomposition.
The zero modes of the 5D fields $W_{L\mu}^{1,2,3}$ and $B_{\mu}$ are identified with the SM gauge bosons.
The fields with $(-+)$ boundary condition only have massive KK modes, the mass of the first KK gauge bosons is of the order $\pi ke^{-kL}$, and as usual it is around 3 TeV within the reach of LHC.
Furthermore, the gauge group $SU(2)_L\times U(1)_Y$ is broken down to $U(1)_{\text{EM}}$ by the VEV of the Higgs localized on the IR brane.  
The Higgs field is a $SU(2)_L\otimes SU(2)_{R}$ bi-doublet and it obtains the following vacuum expectation value,
\begin{equation}
\langle H(x^\mu)\rangle =
\left(\begin{array}{cc}
v & 0  \\ 0 & v
\end{array}\right)\,,
\end{equation}
with $v=174$GeV.
The SM-like neutral electroweak gauge bosons are defined in the usual way $Z_{\mu}=(g_LW^{3}_{L\mu}-g_YB_{\mu})/\sqrt{g^2_L+g^2_Y}$, $A_{\mu}=(g_YW^{3}_{L\mu}+g_LB_{\mu})/\sqrt{g^2_L+g^2_Y}$.
  The $Z$ boson and photon arise as the zero modes of $Z_{\mu}$ and $A_{\mu}$ respectively.
For the family symmetry we choose the $T'$ group.
The $T'$ flavor symmetry has been studied in the literature~\cite{Aranda:1999kc,Aranda:2000tm,Feruglio:2007uu,Chen:2007afa,Ding:2008rj,Frampton:2008bz,Chen:2009gy,Lavoura:2012cv,Meroni:2012ty,Girardi:2013sza}.
We introduce four flavon fields in our model. The flavons $\varphi_{\nu}$ and $\rho_{\nu}$ are localized on the UV brane, while the flavons $\varphi_{l}$ and $\sigma_{l}$ are localized on the IR brane.
  The fermion fields live in the bulk, and the profiles of their zero modes in the fifth dimension are displayed in Fig.~\ref{fig:wavef_lepton}.

\section{Lepton sector}
\label{Sec:lepton}

The transformation properties of the lepton and scalar fields under the $SU(2)_L\times SU(2)_R\times U(1)_{B-L}$ gauge symmetry and $T'\times Z_3\times Z_4$ flavor symmetry are summarized in table~\ref{Tab:assignment_lepton}.
The zero mode of $\Psi_{L}$ is the left-handed lepton doublet, and the zero modes of $\Psi_{e, \mu,\tau}$ and $\Psi_{\nu}$ are the right-handed charged leptons and neutrinos, respectively.
  Their bulk masses are given by $c_{\ell}$, $c_{e,\mu,\tau}$ and $c_{\nu}$, respectively, in units of the AdS curvature.
The vacuum expectation values (VEVs) of the flavon fields are
\begin{table}[hptb!]
\centering
\resizebox{1.0\textwidth}{!}{
\begin{tabular}{|c|c|c|c|c|c|c|c|c|c|c|}
  \hline  \hline
  Field & $\Psi_{l}$ & $\Psi_{e}$ & $\Psi_{\mu}$ & $\Psi_{\tau}$ & $\Psi_{\nu}$ & $H$ & $\varphi_{l}(IR)$ & $\sigma_{l}(IR)$ & $\varphi_{\nu}(UV)$ & $\rho_{\nu}(UV)$  \\
  \hline
  $SU(2)_L\times SU(2)_R\times U(1)_{B-L}$ & $(2,1,-1)$ & $(1,2,-1)$ & $(1,2,-1)$ & $(1,2,-1)$ & $(1,2,-1)$ & $(2,2,0)$ & $(1,1,0)$ & $(1,1,0)$ & $(1,1,0)$ & $(1,1,0)$  \\
  \hline
  $T'$ & $\mathbf{3}$ & $\mathbf{1'}$ & $\mathbf{1''}$ & $\mathbf{1}$ & $\mathbf{3}$ & $\mathbf{1}$ &$\mathbf{2}$ & $\mathbf{1''}$ & $\mathbf{3}$ & $\mathbf{3}$   \\
  \hline
  $Z_{3}$ & $\omega^{2}$ & $1$ &  $1$ &  $1$ & $\omega^{2}$ & $1$ &  $\omega$ &  $\omega$ &  $\omega$ &  $\omega$ \\
\hline
  $Z_{4}$ & $i$ & $i$ &  $i$ &  $i$ & $i$ & $1$ &  $-1$ & $-1$ &  $i$  &  $-i$  \\
\hline \hline
\end{tabular}}
\caption{\label{Tab:assignment_lepton}
  The transformation properties of the lepton sector under the $SU(2)_{L}\times SU(2)_{R}\times U(1)_{B-L}$ gauge group and the $T'\times Z_3\times Z_4$ flavor symmetry, with $\omega = e^{2\pi i/3}$.
  The flavons $\varphi_l$, $\sigma_l$ and $\varphi_{\nu}$, $\rho_{\nu}$ are localized on the IR and UV branes, as indicated.}
\end{table}
\begin{equation}
  \label{eq:vacuum_lepton}
  \braket{\varphi_{l}}=(1,0)v_{\varphi_{l}}\,,\quad
  \braket{\sigma_{l}}=v_{\sigma_{l}}\,,\quad
  \braket{\varphi_{\nu}}=(1, -2\omega^2, -2\omega)v_{\varphi_{\nu}}\,,\quad
  \braket{\rho_{\nu}}=(1, -2\omega, -2\omega^{2})v_{\rho_{\nu}}\,,
\end{equation}
where $\omega = e^{\frac{2i\pi}{3}}$, $v_{\varphi_{l}}$, $v_{\sigma_{l}}$, $v_{\varphi_{\nu}}$ and $v_{\rho_{\nu}}$ are arbitrary complex numbers.
As shown in Appendix~\ref{sec:vacuum_alignment}, the alignment in Eq.~\eqref{eq:vacuum_lepton} is the minimum of the scalar potential.

The leading order charged lepton Yukawa interactions respecting both gauge and flavor symmetries are of the following form,
\begin{equation}
\label{eq:potential_charged_lepton}
\mathcal{L}^{l}_{Y}=\frac{\sqrt{G}}{\Lambda'^3}\Big[
 y _{e}(\varphi_{l}^{2}\overline{\Psi}_{l})_{\mathbf{1''}}H\Psi_{e}
+y_{\mu}(\varphi_{l}^{2}\overline{\Psi}_{l})_{\mathbf{1'}}H\Psi_{\mu}
+y_{\tau}(\varphi_{l}^2\overline{\Psi}_{l})_{\mathbf{1}}H\Psi_{\tau}\Big]
\delta(y-L)+\text{h.c.}\,,
\end{equation}
where $G=e^{-8ky}$ is the determinant of the 5D metric. Inserting the vacuum configuration of Eq.~\eqref{eq:vacuum_lepton} into Eq.~\eqref{eq:potential_charged_lepton} and noticing that 
\begin{equation}
\braket{\varphi_{l}\varphi_{l}}_{\mathbf{3}}=(0,0,1)v^2_{\varphi_{l}}\,,
\end{equation}
then one can read out the charged lepton mass matrix in the zero mode approximation as
\begin{equation}
\label{eq:ml}m_{l}=\frac{1}{\Lambda'^2}v\left( \begin{array}{ccc} \tilde{y}_{e}v_{\varphi_{l}}^2 & 0 & 0 \\ 0 & \tilde{y}_{\mu}v_{\varphi_{l}}^{2} & 0 \\ 0 & 0 & \tilde{y}_{\tau}v_{\varphi_{l}}^2 \\ \end{array} \right)\,,
\end{equation}
where $v$ is the vacuum expectation value of the Higgs field and
\begin{eqnarray}
\tilde{y}_{e,\mu,\tau} = \frac{y_{e,\mu,\tau}}{L\Lambda'}f_L(L,c_\ell)f_R(L,c_{e, \mu, \tau})\,.
\end{eqnarray}
Here $f_{L,R}$ are the zero-mode wave functions of fermion fields, their explicit forms are given in Appendix~\ref{sec:XD}.
One sees that the charged lepton mass matrix is diagonal with
\begin{equation}
m_{e}=\tilde{y}_{e}\frac{v_{\varphi_{l}}^2}{\Lambda'^2}v\,,\quad
m_{\mu}=\tilde{y}_{\mu}\frac{v_{\varphi_{l}}^{2}}{\Lambda'^2}v\,,\quad
m_{\tau}=\tilde{y}_{\tau}\frac{v_{\varphi_{l}}^2}{\Lambda'^2}v\,.
\end{equation}
The correct values of $m_{e, \mu, \tau}$ can be naturally achieved via the wave function overlaps in the usual way.
In our model, neutrino masses are generated by the type-I seesaw mechanism. The corresponding terms invariant under the flavor symmetry $T'\times Z_3\times Z_4$ are given by
\begin{eqnarray}
\nonumber
 \mathcal{L}^{\nu}_{Y}&=&
  y_{\nu_{1}}\frac{\sqrt{G}}{\Lambda'}(\overline{\Psi}_{l}H\Psi_{\nu})_{\mathbf{1}}\delta(y-L) +
  \frac{1}{2}\frac{\sqrt{G}}{\Lambda^{2}}\Big[y_{\nu_{2}}(\overline{N^C}N)_{\mathbf{1}}(\varphi_{\nu}^{2})_{\mathbf{1}}+ y_{\nu_{3}}(\overline{N^C}N)_{\mathbf{1}} (\rho_{\nu}^{2})_{\mathbf{1}} \\
&~&\label{eq:potential_neutrino}
 +y_{\nu_{4}}\left((\overline{N^C}N)_{\mathbf{3_{S}}}(\varphi_{\nu}^{2})_{\mathbf{3_{S}}}\right)_{\mathbf{1}} + y_{\nu_{5}}\left((\overline{N^C}N)_{\mathbf{3_{S}}}(\rho_{\nu}^{2})_{\mathbf{3_{S}}}\right)_{\mathbf{1}}\Big]\delta(y)+\text{h.c.}\,,
\end{eqnarray}
Here $N$ is the $SU(2)_R$ doublet partner of the charged lepton $\tilde{e}$ shown in Eq.~\eqref{eq:leptons-5D} and it is neutral under the SM gauge group with $N^C=C\overline{N}^{T}$,
where $C$ is the charge conjugation matrix (we could also use directly the more fundamental two-component spinor formalism~\cite{Schechter:1980gr}).
Notice that the bulk gauge symmetry $SU(2)_L\times SU(2)_R\times U(1)_{B-L}$ is broken down to the SM gauge group $SU(2)_L\times U(1)_Y$. Thus the lepton number is broken only on the UV brane, and it is
preserved in the bulk and on the TeV brane. As a consequence, a UV brane-localized Majorana mass term for the the right-handed neutrinos $N$ is allowed, as shown in Eq.~\eqref{eq:potential_neutrino}.
Given the vacuum aligment of $\sigma_{\nu}$ and $\varphi_{\nu}$ in Eq.~\eqref{eq:vacuum_lepton}, we can read out the Dirac and Majorana neutrino mass matrices as follows
\begin{eqnarray}
\nonumber  m_{D}&=&\tilde{y}_{\nu_{1}}v\left(\begin{array}{ccc} 1 & 0 & 0 \\ 0 & 1 & 0 \\ 0 & 0 & 1 \\ \end{array} \right)\,,\\
\label{eq:mD-mN}m_{N}&=&(\tilde{y}_{\nu_{2}}\frac{v_{\varphi_{\nu}}^{2}}{\Lambda}+\tilde{y}_{\nu_{3}}\frac{v_{\rho_{\nu}}^{2}}{\Lambda})\left(\begin{array}{ccc} 1 & 0 & 0 \\ 0 & 0 & 1 \\ 0 & 1 & 0 \\ \end{array} \right)+
  \tilde{y}_{\nu_{4}}\frac{v_{\varphi_{\nu}}^{2}}{\Lambda}\left( \begin{array}{ccc} 2 & 2\omega & 2\omega^{2} \\ 2\omega & -4\omega^{2} & -1 \\ 2\omega^{2} & -1 & -4\omega \\ \end{array} \right)+
  \tilde{y}_{\nu_{5}}\frac{v_{\rho_{\nu}}^{2}}{\Lambda}\left( \begin{array}{ccc} 2 & 2\omega^{2} & 2\omega \\ 2\omega^{2} & -4\omega & -1 \\ 2\omega & -1 & -4\omega^{2} \\ \end{array} \right)\,,
\end{eqnarray}
with
\begin{equation}
\tilde{y}_{\nu_{1}} = \frac{y_{\nu_1}}{L\Lambda'} f_L(L,c_\ell)f_R(L,c_{\nu})\,,\qquad
\tilde{y}_{\nu_{2,3,4,5}} = \frac{y_{\nu_{2,3,4,5}}}{L\Lambda}f^2_R(0,c_{\nu})\,.
\end{equation}
Notice that the form of the Yukawa interactions imply that both charged lepton as well as the Dirac neutrino mass blocks are flavor-diagonal.
  Therefore the non-trivial mixing and CP violation parameters required in the physical neutrino mixing matrix must emerge from the type-I seesaw
  mechanism at the scale $m_N$.

By performing the seesaw diagonalization procedure~\cite{Schechter:1981cv}, one gets the effective light neutrino mass matrix expressed in the usual way as
\begin{eqnarray}
m_{\nu}&=&-m_{D}m_{N}^{-1}m_{D}^{T}\\
\nonumber
&=&
m_0\left( \begin{array}{ccc}
\frac{1-2y_{4}-2y_{5}-15y_{4}^{2}+18y_{4}y_{5}-15y_{5}^{2}  }{(3(y_{4}+y_{5})+1)(18(y_{4}-y_{5})^{2}+3(y_{4}+y_{5})-1)}&
\frac{-2\omega(\omega y_{4} + y_{5} + 3\omega y_{4}^{2} + 9\omega^{2} y_{4} y_{5} + 3 y_{5}^{2})}{(3(y_{4}+y_{5})+1)(18(y_{4}-y_{5})^{2}+3(y_{4}+y_{5})-1)}&
\frac{-2\omega(y_{4}+\omega y_{5} + 3 y_{4}^{2} + 9\omega^{2} y_{4} y_{5} + 3\omega y_{5}^{2})}{(3(y_{4}+y_{5})+1)(18(y_{4}-y_{5})^{2}+3(y_{4}+y_{5})-1)}\\
\frac{-2\omega(\omega y_{4} + y_{5} + 3\omega y_{4}^{2} + 9\omega^{2} y_{4} y_{5} + 3 y_{5}^{2})}{(3(y_{4}+y_{5})+1)(18(y_{4}-y_{5})^{2}+3(y_{4}+y_{5})-1)} &
\frac{4(\omega y_{4} + \omega^{2} y_{5} + 3\omega y_{4}^{2} + 3\omega^{2} y_{5}^{2})}{(3(y_{4}+y_{5})+1)(18(y_{4}-y_{5})^{2}+3(y_{4}+y_{5})-1)}&
\frac{1 + y_{4} + y_{5} -6 y_{4}^{2} - 6 y_{5}^{2} }{(3(y_{4}+y_{5})+1)(18(y_{4}-y_{5})^{2}+3(y_{4}+y_{5})-1)}\\
\frac{-2\omega(y_{4}+\omega y_{5} + 3 y_{4}^{2} + 9\omega^{2} y_{4} y_{5} + 3\omega y_{5}^{2})}{(3(y_{4}+y_{5})+1)(18(y_{4}-y_{5})^{2}+3(y_{4}+y_{5})-1)} &
\frac{1 + y_{4} + y_{5} -6 y_{4}^{2} - 6 y_{5}^{2} }{(3(y_{4}+y_{5})+1)(18(y_{4}-y_{5})^{2}+3(y_{4}+y_{5})-1)} &
\frac{4(\omega^{2} y_{4} + \omega y_{5} + 3\omega^{2} y_{4}^{2} + 3\omega y_{5}^{2})}{(3(y_{4}+y_{5})+1)(18(y_{4}-y_{5})^{2}+3(y_{4}+y_{5})-1)}\\
\end{array} \right)\,,
\end{eqnarray}
where $m_0=\frac{\tilde{y}_{\nu_{1}}^{2}\Lambda v^{2}}{\tilde{y}_{\nu_{2}}v_{\varphi_{\nu}}^{2}+\tilde{y}_{\nu_{3}}v_{\rho_{\nu}}^{2}}$, $y_{4}=\frac{\tilde{y}_{\nu_{4}}v_{\varphi_{\nu}}^{2}}{\tilde{y}_{\nu_{2}}v_{\varphi_{\nu}}^{2}+\tilde{y}_{\nu_{3}}v_{\rho_{\nu}}^{2}}$ and $y_{5}=\frac{\tilde{y}_{\nu_{5}}v_{\rho_{\nu}}^{2}}{\tilde{y}_{\nu_{2}}v_{\varphi_{\nu}}^{2}+\tilde{y}_{\nu_{3}}v_{\rho_{\nu}}^{2}}$.
It is remarkable that, apart from an overall mass scale $m_0$, the mass matrix $m_{\nu}$ only depends on two complex input parameters $y_4$, $y_5$. These will describe the three neutrino masses and also lead to predictions for the lepton mixing matrix. We first perform a tri-bimaximal transformation on the neutrino fields. The resulting light neutrino mass matrix becomes
\begin{eqnarray}
\nonumber
  m_{\nu}'&=&U_{TBM}^{\dagger}m_{\nu}U_{TBM}^{*}\\
  &=&m_0
\left( \begin{array}{ccc}
\frac{-1}{1+3(y_{4}+y_{5})} & 0 & 0 \\
0 & \frac{1-3(y_{4} + y_{5})}{18(y_{4}-y_{5})^{2}+3(y_{4}+y_{5})-1} & \frac{-3\sqrt{2}i(y_{4}-y_{5})}{18(y_{4}-y_{5})^{2}+3(y_{4}+y_{5})-1} \\
0 & \frac{-3\sqrt{2}i(y_{4}-y_{5})}{18(y_{4}-y_{5})^{2}+3(y_{4}+y_{5})-1} & \frac{-1}{18(y_{4}-y_{5})^{2}+3(y_{4}+y_{5})-1}
\end{array} \right)\,,
\end{eqnarray}
where $U_{TBM}$ is the well-known tri-bimaximal mixing matrix,
\begin{equation}
  U_{TBM}=\left( \begin{array}{ccc} \sqrt{\frac{2}{3}} & \frac{1}{\sqrt{3}} & 0 \\
                   -\frac{1}{\sqrt{6}} & \frac{1}{\sqrt{3}} & -\frac{1}{\sqrt{2}} \\
                   -\frac{1}{\sqrt{6}} & \frac{1}{\sqrt{3}} & \frac{1}{\sqrt{2}} \\ \end{array} \right)\,.
\end{equation}
Since $m_{\nu}'$ is a block-diagonal symmetric matrix, it can be exactly diagonalized as
\begin{equation}
  U_{\nu}'^{\dagger}m_{\nu}'U_{\nu}'^{*}=\text{diag}(m_{1},m_{2},m_{3})\,,
\end{equation}
where $U_{\nu}'$ can be generally denoted as
\begin{equation}
  U_{\nu}'=\left(\begin{array}{ccc} 1 & 0 & 0 \\ 0 & \cos\theta_{\nu} & \sin\theta_{\nu} e^{i\delta_{\nu}} \\ 0 & -\sin\theta_{\nu} e^{-i\delta_{\nu}} & \cos\theta_{\nu} \\ \end{array} \right)\,.
\end{equation}
Since the charged lepton mass matrix $m_{l}$ is diagonal in this case, the lepton mixing matrix is determined to be\footnote{We notice that the first column of the lepton mixing matrix is fixed to be $(2, -1,-1)^{T}/\sqrt{6}$.}
\begin{eqnarray}
\nonumber  U&=&U_{TBM}U_{\nu}'\\
\nonumber  &=&  \left( \begin{array}{ccc} \sqrt{\frac{2}{3}} & \frac{1}{\sqrt{3}} & 0 \\
-\frac{1}{\sqrt{6}} & \frac{1}{\sqrt{3}} & -\frac{1}{\sqrt{2}} \\
-\frac{1}{\sqrt{6}} & \frac{1}{\sqrt{3}} & \frac{1}{\sqrt{2}} \\ \end{array} \right).\left(\begin{array}{ccc} 1 & 0 & 0 \\ 0 & \cos\theta_{\nu} & \sin\theta_{\nu} e^{i\delta_{\nu}} \\ 0 & -\sin\theta_{\nu} e^{-i\delta_{\nu}} & \cos\theta_{\nu} \\ \end{array} \right)\,,\\
\label{eq:Umix}&=&
\left( \begin{array}{ccc}
\sqrt{\frac{2}{3}} &
\frac{\cos\theta_{\nu}}{\sqrt{3}} &
\frac{\sin\theta_{\nu} e^{i\delta_{\nu}}}{\sqrt{3}} \\
-\frac{1}{\sqrt{6}} &
\frac{\cos\theta_{\nu}}{\sqrt{3}} + \frac{\sin\theta_{\nu} e^{-i\delta_{\nu}}}{\sqrt{2}} &
-\frac{\cos\theta_{\nu}}{\sqrt{2}} + \frac{\sin\theta_{\nu} e^{i\delta_{\nu}}}{\sqrt{3}} \\
-\frac{1}{\sqrt{6}} &
\frac{\cos\theta_{\nu}}{\sqrt{3}} - \frac{\sin\theta_{\nu} e^{-i\delta_{\nu}}}{\sqrt{2}} &
\frac{\cos\theta_{\nu}}{\sqrt{2}} + \frac{\sin\theta_{\nu} e^{i\delta_{\nu}}}{\sqrt{3}}\\ \end{array} \right)\,.
\end{eqnarray}

\subsection{Predictions for neutrino oscillations}
\label{sec:neutr-oscill}

We start this section by noticing that, in the absence of the Majorana terms in Eq.~\eqref{eq:potential_neutrino}, in this model neutrinos would be unmixed,
since both charged lepton and Dirac mass terms are simultaneously diagonal.
 They would also be degenerate in mass.
 Hence the neutrino mass differences, as well as mixing and CP violation parameters, all result from the seesaw mechanism.
This is in sharp contrast with the warped standard model extension proposed in Ref.~\cite{Chen:2015jta}.\\[-.2cm]

From the lepton mixing matrix obtained in Eq.~\eqref{eq:Umix}, one can easily extract the following results for the neutrino mixing angles as well as the leptonic Jarlskog invariant,
\begin{eqnarray}
&&\sin^{2}\theta_{13}=\frac{\sin^{2}\theta_{\nu}}{3}\,,\\
&&\sin^{2}\theta_{12}=1-\frac{4}{5+\cos2\theta_{\nu}}\,,\\
&&\sin^{2}\theta_{23}=\frac{1}{2}-\frac{\sqrt{6}\sin2\theta_{\nu}\cos\delta_{\nu}}{5+\cos2\theta_{\nu}}\,,\\
&&J_{CP}=\frac{\sin2\theta_{\nu}\sin\delta_{\nu}}{6\sqrt{6}}\,.
\end{eqnarray}
One sees that the three neutrino mixing angles as well as the Dirac CP violation phase are all expressed in terms of just two parameters, $\theta_{\nu}$ and $\delta_{\nu}$.
Therefore there are two relations between these mixing angles and the Dirac CP violation phase, that can be expressed analytically as
\begin{equation}
\label{eq:osc-predictions}
\cos^{2}\theta_{12}\cos^{2}\theta_{13}=\frac{2}{3}\,,\quad \cos\delta_{CP}=\frac{(3\cos 2\theta_{12}-2)\cos 2\theta_{23}}{3\sin 2\theta_{23}\sin2\theta_{12}\sin\theta_{13}}\,.
\end{equation}
\begin{figure}[h!]
\centering
\begin{tabular}{cc}
\includegraphics[width=0.48\linewidth]{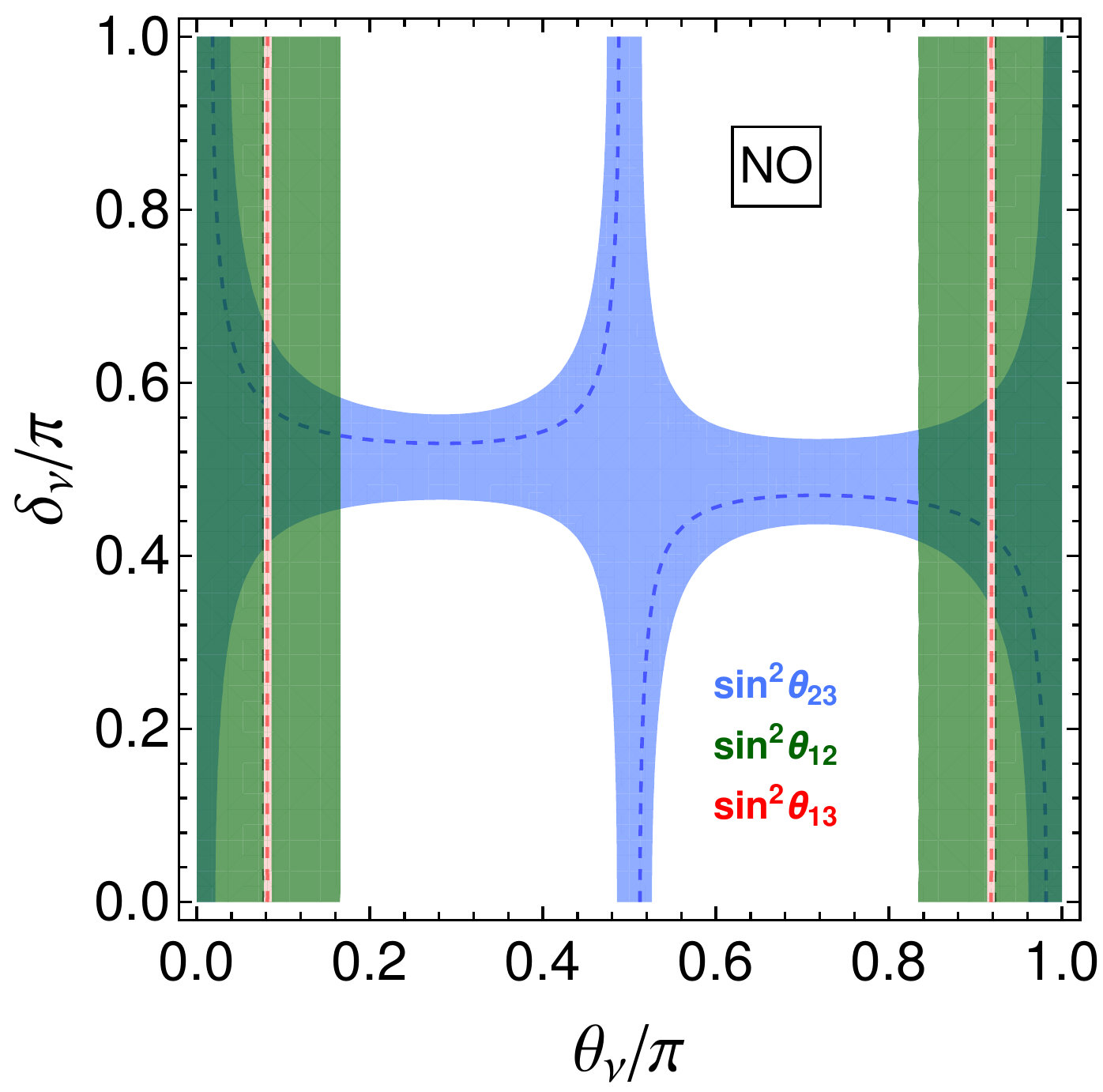}
\includegraphics[width=0.48\linewidth]{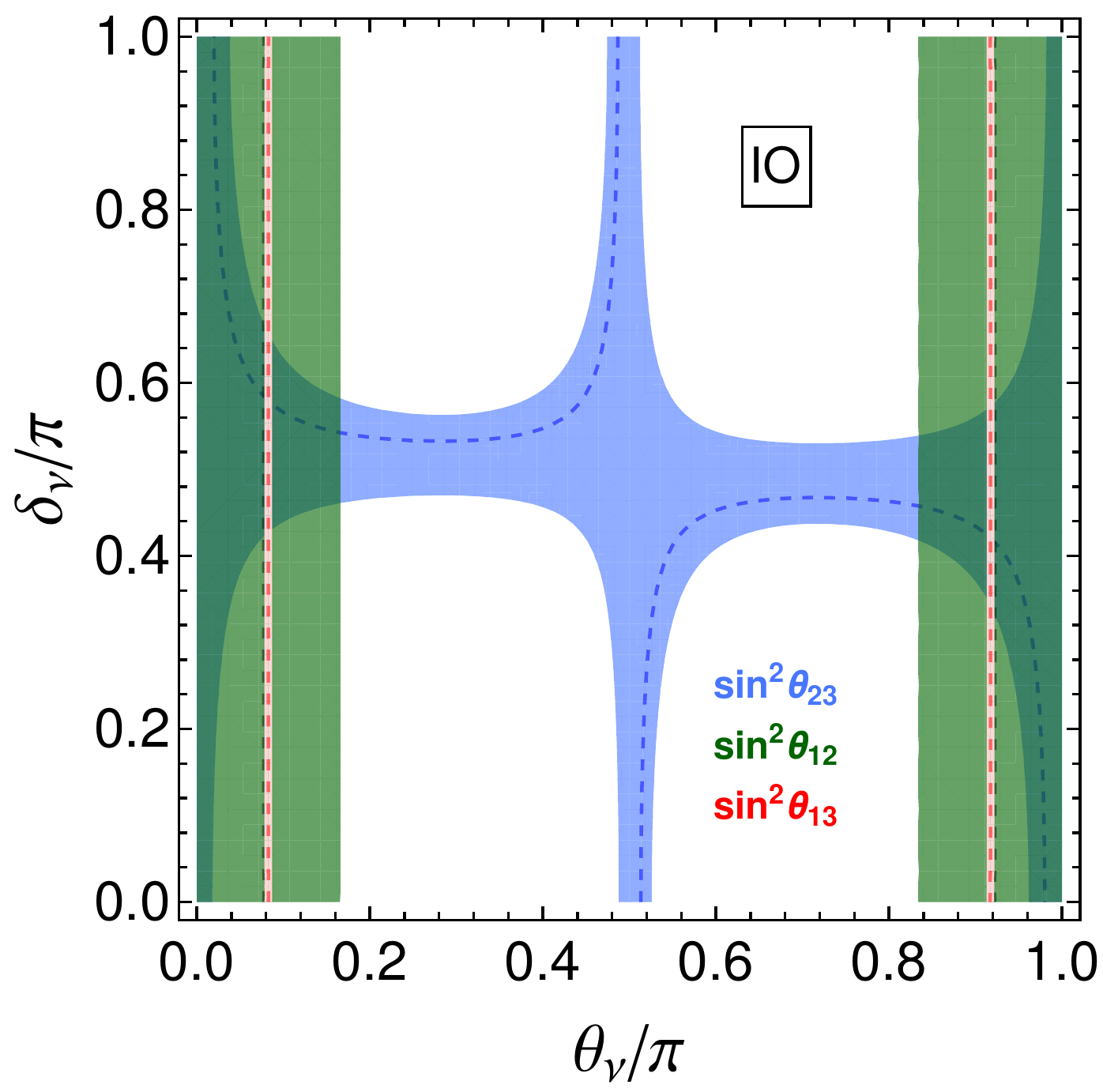}\\
\end{tabular}
\caption{\label{fig:osc-predictions-I} Contour plots of $\sin^2\theta_{12}$, $\sin^2\theta_{13}$, and $\sin^{2}\theta_{23}$ in the $\theta_{\nu}-\delta_{\nu}$ plane. The red, green and blue areas denote the $3\sigma$ regions of $\sin^{2}\theta_{13}$, $\sin^{2}\theta_{12}$ and $\sin^{2}\theta_{23}$ respectively, and the dashed lines refer to their best fit values taken from \cite{deSalas:2017kay}.}
\end{figure}
\begin{figure}[h!]
\centering
\begin{tabular}{cc}
\includegraphics[width=0.48\linewidth]{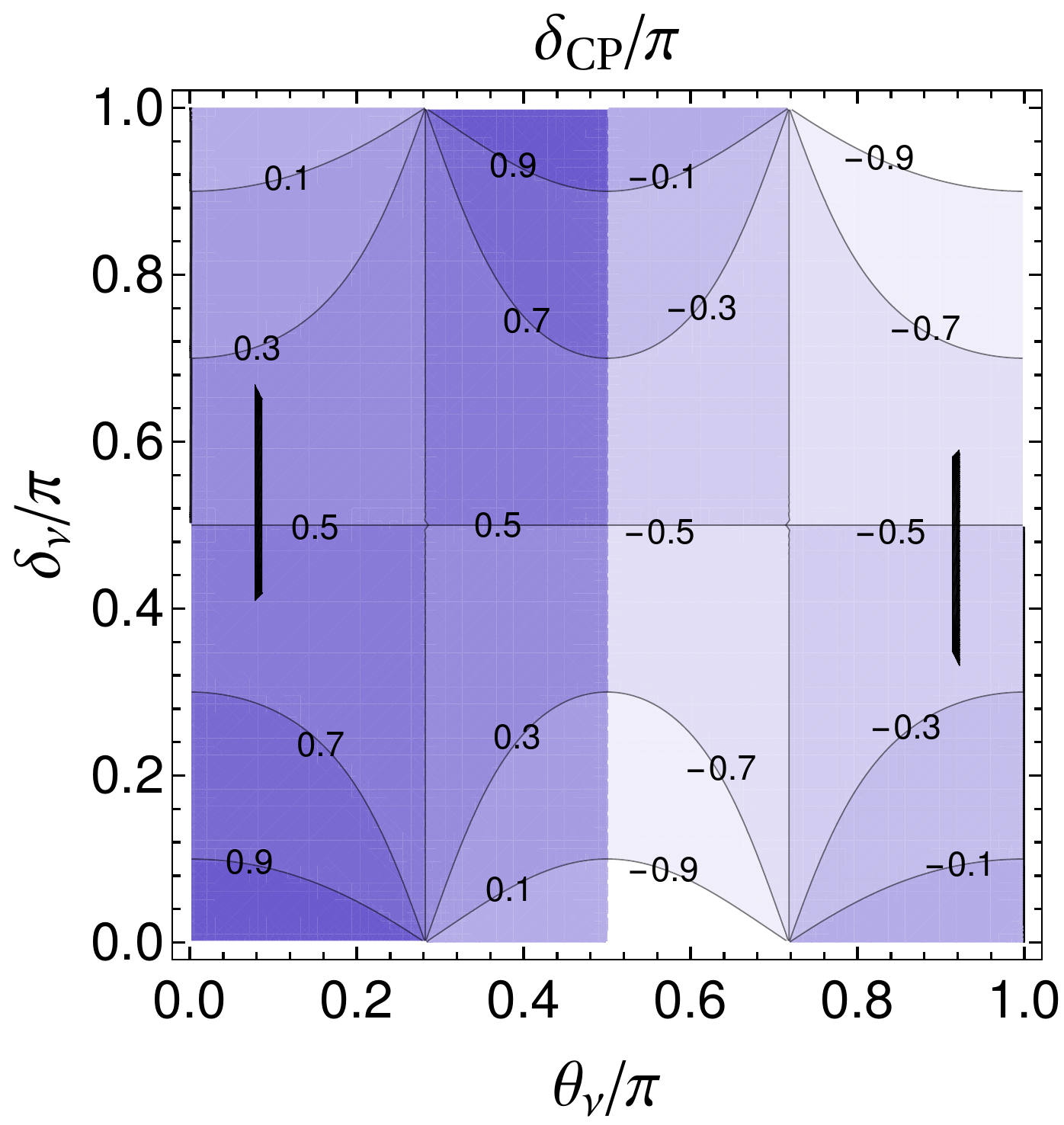}
\includegraphics[width=0.48\linewidth]{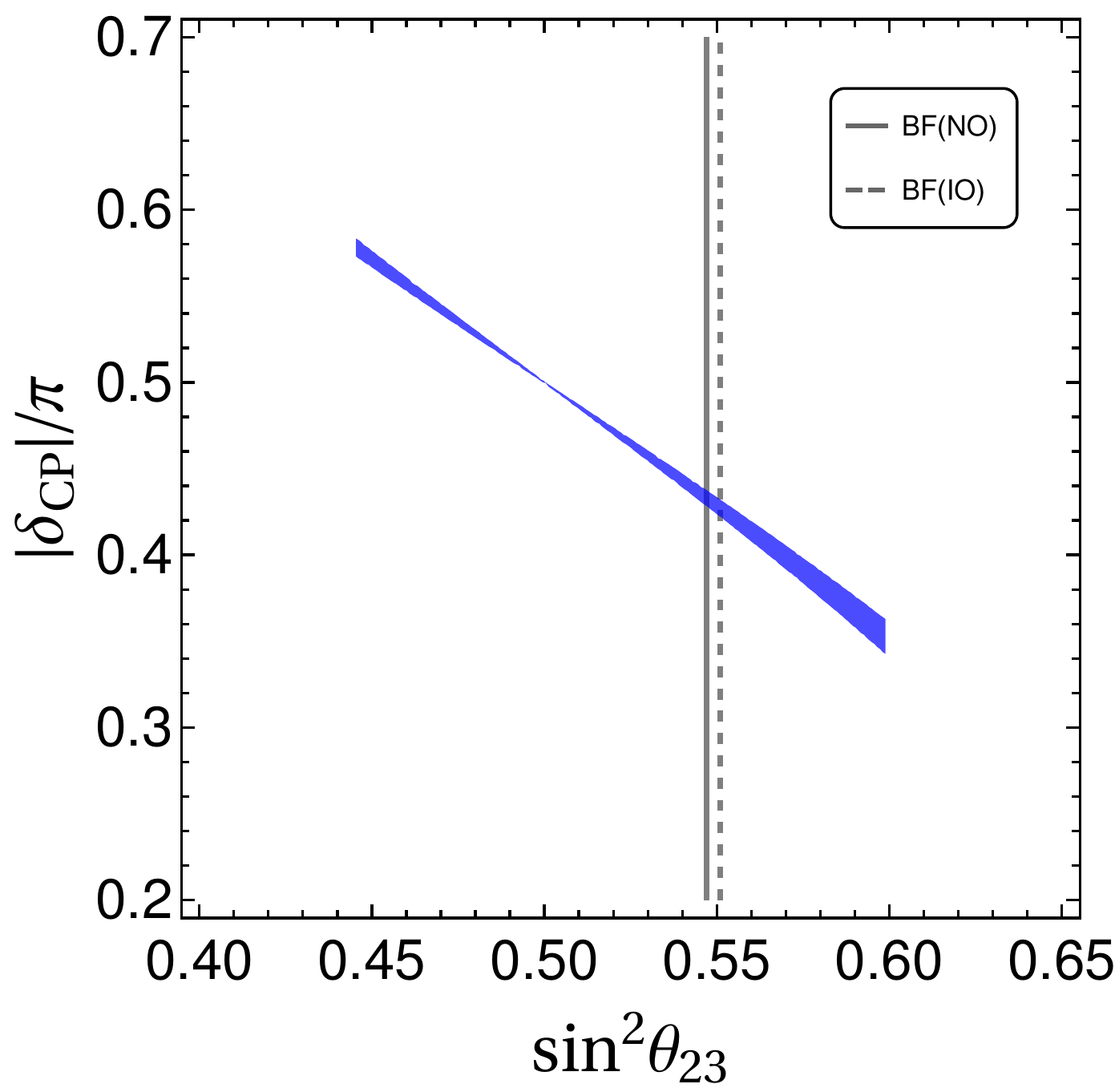}
\end{tabular}
\caption{\label{fig:osc-predictions-II} Contour plots of $\delta_{CP}$ in the $\theta_{\nu}-\delta_{\nu}$ plane (left) and correlation between $|\delta_{CP}|$ and $\sin^{2}\theta_{23}$ (right).
The black areas correspond to the  $3\sigma$ allowed regions of lepton mixing angles~\cite{deSalas:2017kay}. The vertical solid and dashed lines in the right panel represent the best fit values of $\sin^2\theta_{23}$ for NO and IO, respectively. }
 \end{figure}

In Fig.~\ref{fig:osc-predictions-I} we display the contour plots of $\sin^2\theta_{12}$, $\sin^2\theta_{13}$, $\sin^2\theta_{23}$ and Dirac CP violation phase $\delta_{CP}$ in the $\theta_{\nu}-\delta_{\nu}$ plane.
The shaded regions are the ones allowed by individual measurements of the three mixing angles, according to the global oscillation analysis in Ref.~\cite{deSalas:2017kay}.
One sees that the parameter $\theta_{\nu}$ is constrained to lie within quite narrow regions around $\theta_{\nu}\simeq0.082\pi$ and $\theta_{\nu}\simeq0.918\pi$.
The left panel in Fig.~\ref{fig:osc-predictions-II} shows the contour plots of $\delta_{CP}$ in the $\theta_{\nu}-\delta_{\nu}$ plane. The black bands denote the regions in which all the three lepton mixing angles lie in the experimentally allowed $3\sigma$ ranges~\cite{deSalas:2017kay}. As in most flavor models, for example those based on modular symmetries, the sign of $\delta_{CP}$ can not be fixed uniquely, the predicted correlation between $|\delta_{CP}|$ and $\sin^2\theta_{23}$ is shown in the right panel of Fig.~\ref{fig:osc-predictions-II}.

\subsection{Predictions for the absolute neutrino mass scale}
\label{sec:neutr-double-beta}

As we already saw, in our model the neutrino mass differences as well as mixing and CP violation all result from the seesaw mechanism.
This is in sharp contrast with the warped standard model extension proposed in Ref.~\cite{Chen:2015jta}.
This implies that in our present model neutrinos must be Majorana particles, leading to the existence of neutrinoless double beta decay, or \znbb for short.

\begin{figure}[h!]
\centering
\begin{tabular}{c}
\includegraphics[width=0.5\linewidth]{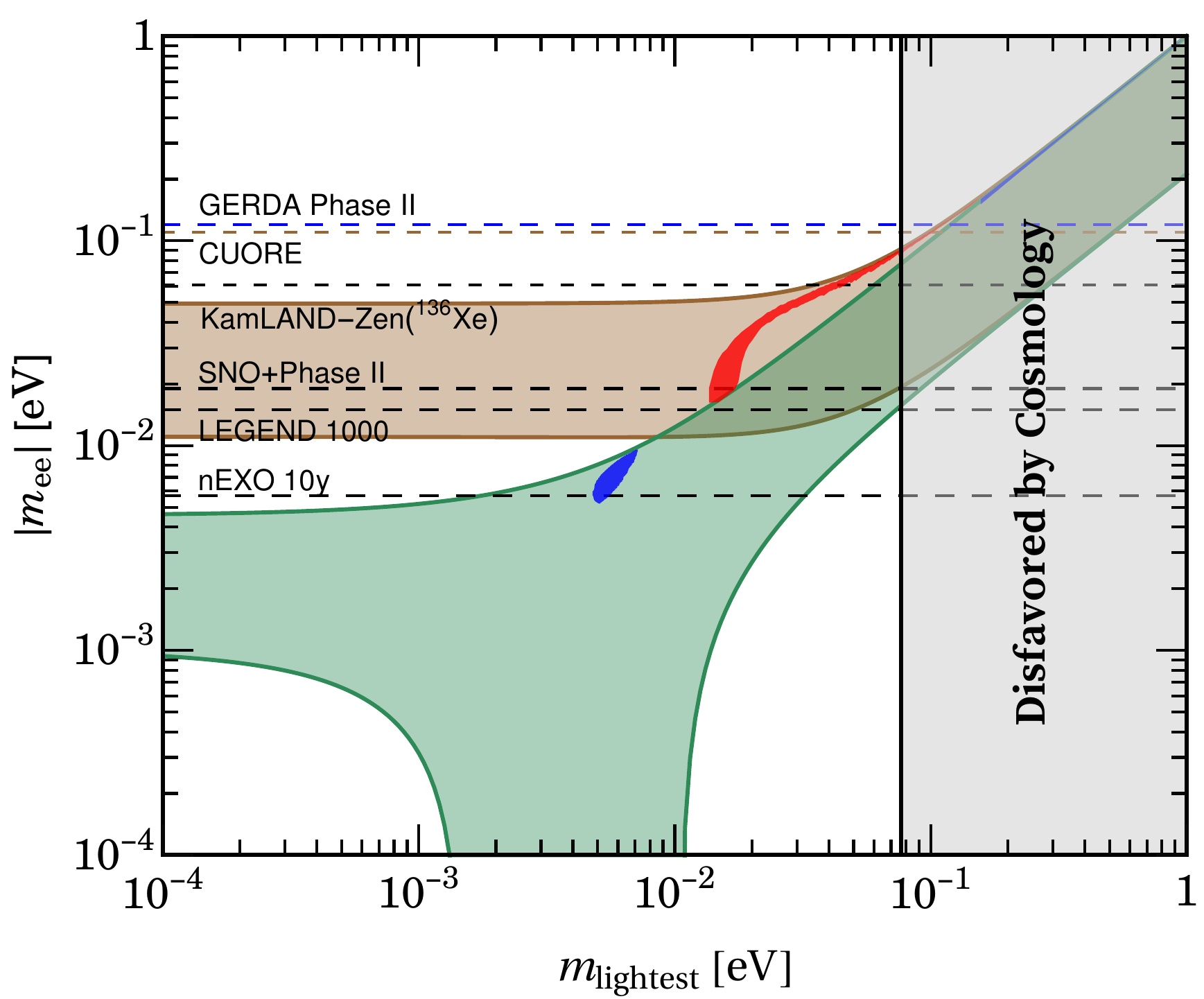}\\
\end{tabular}
 \caption{\label{fig:znbb} Expected mass parameter characterizing the \znbb amplitude, where the red and blue regions are for IO and NO, respectively. The values of the neutrino oscillation parameters are taken from \cite{deSalas:2017kay}. The vertical grey exclusion band denotes the current bound coming from the cosmological data of $\Sigma_{i}m_{i}<0.120\text{eV}$ at $95\%$ confidence level obtained by the Planck collaboration~\cite{Aghanim:2018eyx}.}
\end{figure}

One can determine the expected ranges for the \znbb decay amplitude, taking into account the allowed neutrino oscillation parameters obtained from experiment~\cite{deSalas:2017kay}.
In Fig.~\ref{fig:znbb} we plot the expected values for the mass parameter $|m_{ee}|$ characterizing the \znbb amplitude.
In a generic model the regions expected for inverted-ordered and normal-ordered neutrino masses are indicated by the broad shaded regions indicated in Fig.~\ref{fig:znbb}.

The current experimental bound from KamLAND-Zen~\cite{KamLAND-Zen:2016pfg} as well as the estimated experimental sensitivities are indicated by the horizontal lines~\cite{Alduino:2017ehq,Albert:2017owj,Agostini:2018tnm,Andringa:2015tza,Abgrall:2017syy,Albert:2017hjq}~\footnote{Note that for all of them we have assumed ``optimistic'' values for the corresponding nuclear matrix elements.}.
We now show how, within our model, the predictions for the oscillation parameters imply important restrictions on the effective Majorana mass $|m_{ee}|$. In fact, the allowed ranges are quite narrow. If the neutrino mass spectrum is inverted-ordered (IO), the effective Majorana masss has a lower limit $|m_{ee}|\geq 0.0162~\text{eV}$, while the lightest neutrino mass satisfies $m_{\text{lightest}}\geq 0.0133~\text{eV}$. In contrast, in the case of normal-ordering (NO), the effective mass $|m_{ee}|$ lies in the narrow interval $[5.2\text{meV}, 9.6\text{meV}]$, and the allowed region of $m_{\text{lightest}}$ is $[4.8\text{meV}, 7.2\text{meV}]$\footnote{As shown in Fig.~\ref{fig:znbb}, the neutrino mass spectrum could possibly be quasi-degenerate as well, however this region is disfavored by both KamLAND-Zen and Planck.}. As indicated in the figure, we expect that these predictions will be tested by the next generation $0\nu\beta\beta$ decay experiments.

The predicted neutrino mass parameters relevant for endpoint $\beta$ decay studies as well as cosmology are also interesting,
  as indicated in table~\ref{Tab:mnu_beta_sum}.
These should be compared with the recent limits from the KATRIN experiment~\cite{Aker:2019uuj}, and the $95\%$ confidence limit for the sum of neutrino masses set by the Planck collaboration~\cite{Aghanim:2018eyx}.

\begin{table}[hptb!]
\centering
\begin{tabular}{|c|c|c|}
\hline
\hline
parameter & Experimental results & Predictions\\
\hline
\hline
$m_{\beta}~[\text{meV}]$ (NO) & \multirow{2}{*}{$<1100$} & 10.62 \\
$m_{\beta}~[\text{meV}]$ (IO) & & 52.07 \\
\hline
$\Sigma_{i}m_{i}~[\text{meV}]$ (NO) & \multirow{2}{*}{$<120$} & 66.81 \\
$\Sigma_{i}m_{i}~[\text{meV}]$ (IO) & & 123.34 \\
\hline
\hline
\end{tabular}
\caption{\label{Tab:mnu_beta_sum}
The predictions for the effective neutrino mass $m_{\beta}$ in $\beta$ decay and the sum of neutrino masses. The latest experimental bounds on $m_{\beta}$ and $\Sigma_{i}m_{i}$ are taken from KATRIN~\cite{Aker:2019uuj} and Planck 2018~\cite{Aghanim:2018eyx}, respectively.}
\end{table}

\section{Quark sector}
\label{sec:quarks}

\begin{table}[b!]
\centering
\resizebox{1.0\textwidth}{!}{
\begin{tabular}{|c|c|c|c|c|c|c|c|c|c|c|}
  \hline  \hline
  Field & $\Psi_{UC}$ & $\Psi_{T}$ & $\Psi_{u}$ & $\Psi_{c}$ & $\Psi_{t}$ & $\Psi_{ds}$ & $\Psi_{b}$ & $H$ & $\varphi_{l}(IR)$ & $\sigma_{l}(IR)$\\
  \hline
  $SU(2)_L\times SU(2)_R\times U(1)_{B-L}$ & $(2,1,1/3)$ & $(2,1,1/3)$ & $(1,2,1/3)$ & $(1,2,1/3)$ & $(1,2,1/3)$ & $(1,2,1/3)$ & $(1,2,1/3)$ & $(2,2,0)$ & $(1,1,0)$ & $(1,1,0)$ \\
  \hline
  $T'$ & $\mathbf{2}$ & $\mathbf{1}$ & $\mathbf{1'}$ & $\mathbf{1''}$ & $\mathbf{1'}$ & $\mathbf{2'}$ &$\mathbf{1''}$ & $\mathbf{1}$ & $\mathbf{2}$ & $\mathbf{1''}$ \\
  \hline
  $Z_{3}$ & $\omega^{2}$ & $\omega$ &  $1$ &  $\omega^{2}$ & $1$ & $\omega$ &  $\omega^{2}$ & $1$ & $\omega$ & $\omega$ \\
\hline
  $Z_{4}$ & $1$ & $-1$ &  $1$ &  $-1$ & $-1$ & $1$ &  $-1$ & $1$ & $-1$ & $-1$  \\
\hline \hline
\end{tabular}}
\caption{\label{Tab:assignment_quark} The transformation properties of the quark fields under the bulk gauge group $SU(2)_{L}\times SU(2)_{R}\times U(1)_{B-L}$ and the flavor symmetry $T'\times Z_3\times Z_4$.  Note that no new scalars are needed beyond those in table~\ref{Tab:assignment_lepton}.}
\end{table}
\begin{figure}[b!]
\centering
\begin{tabular}{cc}
\includegraphics[width=0.5\linewidth]{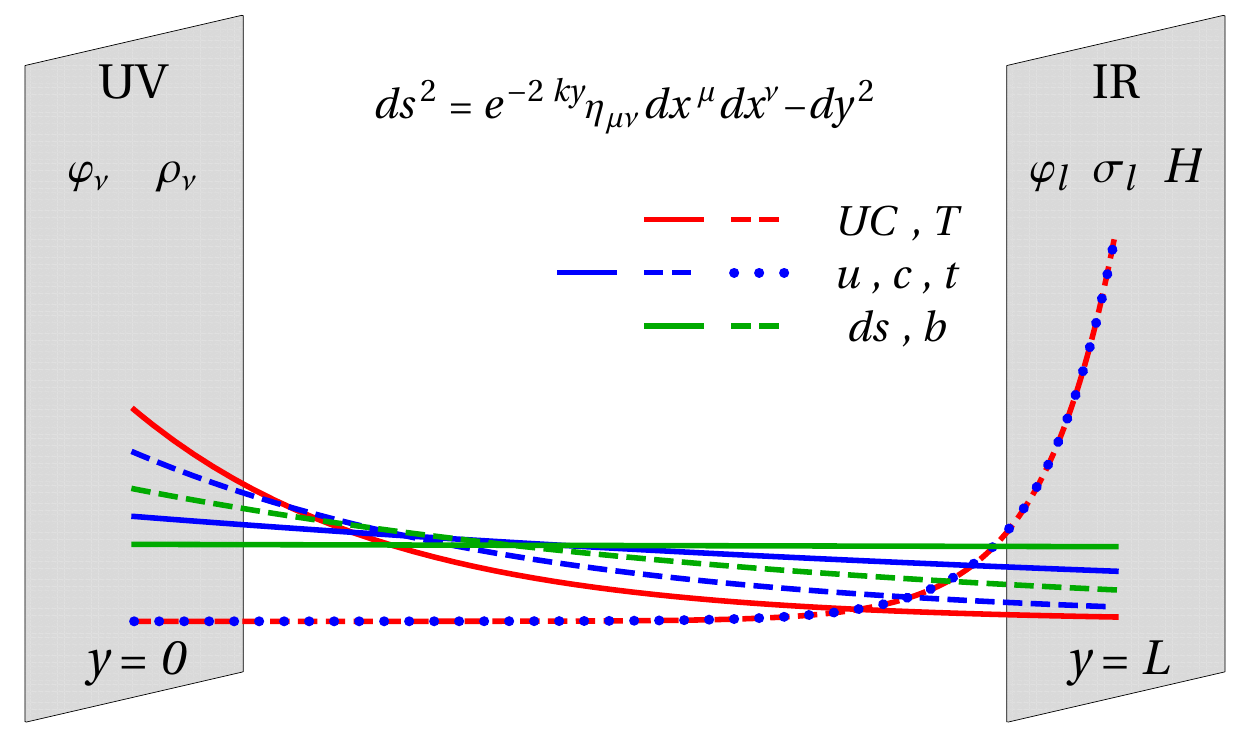}
\end{tabular}
\caption{ The wave functions of the zero modes of the quark fields. }
\label{fig:wavef_quark}
\end{figure}

We now extend our model to the quark sector. The classification of the quark fields under the flavor symmetry $T'\times Z_3\times Z_4$ is given in table~\ref{Tab:assignment_quark}, and no new flavon fields are required. We show the profiles of the zero modes of the quark fields in Fig.~\ref{fig:wavef_quark}. It is straightforward to read off the down-type quark Yukawa interactions
\begin{equation}
\label{eq:LYd}\mathcal{L}^{d}_{Y}=\frac{\sqrt{G}}{\Lambda'^3}\Big[
   y_{ds_1} (\overline{\Psi}_{UC}{H}\Psi_{ds})_{\mathbf{3}} \varphi_{l}^{\ast\,2}
 + y_{ds_2} (\overline{\Psi}_{UC}{H}\Psi_{ds})_{\mathbf{1'}} \sigma_l^{\ast\,2}
 + y_{b}' (\overline{\Psi}_{T}{H}\Psi_{b})_{\mathbf{1''}}  \sigma_l^{2}
\Big]\delta(y-L) + \text{h.c} + \cdots \,,
\end{equation}
where dots stand for higher dimensional operators. Similarly, the up-type quark Yukawa interactions take the form
\begin{eqnarray}
\nonumber
\mathcal{L}^{u}_{Y}&=&\frac{\sqrt{G}}{\Lambda'^3}\Big[
 y_{u}' \Lambda' (\overline{\Psi}_{T}{H}\Psi_{u})_{\mathbf{1'}} \sigma_l
 + y_t \Lambda' (\overline{\Psi}_{UC}{H}\Psi_{t})_{\mathbf{2}} \varphi_{l}^{*}
 + y_u (\overline{\Psi}_{UC}{H}\Psi_{u})_{\mathbf{2}} \varphi_{l}\sigma_l\\
\label{eq:LYu}&&\qquad\qquad
 + y_{c}' (\overline{\Psi}_{T}{H}\Psi_{c})_{\mathbf{1''}} \sigma_{l}^{2}
 + y_t' (\overline{\Psi}_{T}{H}\Psi_{t})_{\mathbf{1'}} \sigma_{l}^{\ast 2}\Big]\delta(y-L)+ \text{h.c}+\cdots\,.
\end{eqnarray}
In the zero mode approximation, we integrate over the fifth dimension and then obtain the up-type and down-type quark mass matrices as follows
\begin{eqnarray}
\label{eq:mass_quark-down}
m^{d}&=&v
\left(\begin{array}{ccc}
  \tilde{y}_{ds_2}v_{\sigma_l}^{\ast2}/\Lambda'^2 & 0 & 0 \\
  \tilde{y}_{ds_1}v_{\varphi_l}^{\ast2}/\Lambda'^2 & \tilde{y}_{ds_2}v_{\sigma_l}^{\ast2}/\Lambda'^2 & 0 \\
0&0 & \tilde{y}_{b}'v_{\sigma_l}^{2}/\Lambda'^2
\end{array}\right)\,,\\
\label{eq:mass_quark-up}m^{u}&=&v
\left(\begin{array}{ccc}
  \tilde{y}_u v_{\varphi_l}v_{\sigma_l}/\Lambda'^2 & 0 & 0 \\
  0 & 0 & \tilde{y}_t v_{\varphi_l}^\ast/\Lambda' \\
  \tilde{y}_u' v_{\sigma_l}/\Lambda' &
  \tilde{y}_c' v_{\sigma_{l}}^{2}/\Lambda'^{2} &
  \tilde{y}_t' v_{\sigma_l}^{\ast2}/\Lambda'^2 \\
\end{array}\right)\,.
\end{eqnarray}
with
\begin{eqnarray}
\tilde{y}_{u,t,ds_{1,2}}=\frac{y_{u,t,ds_{1,2}}}{L\Lambda'}f_L(L,c_{UC})f_R(L,c_{u,t,ds})\,,\qquad
\tilde{y}_{u,c,t,b}^\prime = \frac{y_{u,c,t,b}^\prime}{L\Lambda'}f_L(L,c_{T})f_R(L,c_{u,c,t,b})\,.
\end{eqnarray}
For simplicity, we denote the $ij$ element of $m^u$ ($m^d$) as $m^u_{ij}$ ($m^d_{ij}$). The down-type quark mass matrix is block diagonal with $m^d_{11}=m^d_{22}$, and it can easily diagonalized by a unitary transformation $U_d$,
\begin{equation}
U_{d}=
\left(\begin{array}{ccc}
\cos{\theta_d}                     &\sin{\theta_d} e^{i\varphi_d}             &   0 \\
-\sin{\theta_d} e^{-i\varphi_d}    &\cos{\theta_d}                            &   0 \\
0          &0  &    1
\end{array}\right)\,,
\end{equation}
with
\begin{equation}
\tan{2\theta_d}=|2m^d_{11}/m^{d}_{21}|,~~~\varphi_d=\arg(m^d_{11} m^{d*}_{21})\,.
\end{equation}
The down-type quark masses are determined to be
\begin{equation}
m_{d,s}=\sqrt{|m^d_{11}|^2+|m^d_{21}|^2/2\pm|m^d_{21}|\sqrt{|m^d_{11}|^2+|m^d_{21}|^2/4}}\,,~~~~~ m_b=|m^b_{33}|\,.
\end{equation}
The product of the up-type quark mass matrix with its hermitian conjugate is of the following form
\begin{equation}
m^u m^{u\dagger}=
\left(\begin{array}{ccc}
|m^u_{11}|^2 & 0 & m^u_{11} m^{u\ast}_{31}\\
0 & |m^u_{23}|^2 & m^u_{23} m^{u\ast}_{33}\\
m^{u\ast}_{11}m^u_{31} & m^{u\ast}_{23}m^{u}_{33} & |m^u_{31}|^2 + |m^u_{32}|^2+|m^u_{33}|^2
\end{array}\right)\,.
\end{equation}
The resulting up-type diagonalization matrix can be parameterized as
\begin{equation}
 U_{u}\simeq
\left(\begin{array}{ccc}
1&\epsilon\sin{\theta_u} e^{-i\varphi_u}&-\epsilon\cos{\theta_u}\\
0&\cos{\theta_u}                         &\sin{\theta_u} e^{i\varphi_u}          \\
\epsilon^\ast&-\sin{\theta_u} e^{-i\varphi_u}        &\cos{\theta_u}
\end{array}\right),~~
\end{equation}
where
\begin{equation}
 \tan{2\theta_u}=\frac{2|m^u_{23}m^{u}_{33}|}{|m^u_{33}|^2+|m^u_{32}|^2+|m^u_{31}|^2-|m^u_{23}|^2}\,,~
 \epsilon =\frac{-m^{u}_{11}m^{u\ast}_{31}}{|m^u_{33}|^2+|m^u_{32}|^2+|m^u_{31}|^2-|m^u_{11}|^2},~~\varphi_u=\arg(m^u_{23} m^{u*}_{33})\,.
\end{equation}
We find the up-type quark mass eigenvalues are
\begin{equation}
m_u\simeq|m^u_{11}|\sqrt{1-\frac{|m^u_{23}|^2|m^u_{31}|^2}{m_c^2m_t^2}}\,, \qquad m_{c,t}=\frac{1}{\sqrt{2}}
\sqrt{X^+\pm\sqrt{(X^-)^2+4|m^u_{23}|^2 |m^{u}_{33}|^2}}\,,
\end{equation}
with $X^{\pm}=|m^u_{33}|^2+|m^u_{32}|^2+|m^u_{31}|^2\pm|m^u_{23}|^2$. As a result, the quark mixing matrix is given by
\begin{eqnarray}
V_{\text{CKM}}
&=&U_{u}^\dagger  U_{d}
\\
&\simeq&
\left(
\begin{array}{ccc}
 \cos \theta _d &
 e^{i\varphi _d} \sin \theta_d &
 \epsilon
\\
 -e^{-i \varphi _d} \cos \theta_u \sin \theta_d - e^{i \varphi_u} \sin \theta_u \cos \theta_d \epsilon^*&
 \cos \theta_d \cos \theta_u  - e^{i (\varphi_u + \varphi_d ) } \sin \theta_u \sin \theta_d \epsilon^* &
 -e^{i \varphi _u} \sin \theta _u
\\
 -e^{-i \left(\varphi _d+\varphi_u\right)} \sin \theta_d \sin \theta_u - \cos \theta_u \cos \theta_d \epsilon^*    &
 e^{-i \varphi _u} \cos \theta_d \sin\theta_u -  e^{i \varphi_d} \cos \theta_u \sin \theta_d \epsilon^*  &
 \cos\theta _u
\\
\end{array}
\right)\,,
\nonumber
\end{eqnarray}
from which we can extract the expressions of CP violation phase and Jarlskog invariant in the quark sector as follows,
\begin{eqnarray}
\delta^q_{\text{CP}}&=&\pi-\arg(\epsilon)+\varphi _d+\varphi_u\,,\label{deltaq}\\
J^{q}_{\text{CP}}&\simeq & \frac{1}{4} |\epsilon| \sin 2 \theta _d \sin 2 \theta _u \sin \delta^q_{\text{CP}}\label{Jq}\,.
\end{eqnarray}
Besides, we can find that in this case $\theta_{c}\simeq \theta_{d}$.
With the fact that the down quark mass matrix is block-digonalized and it satisfies the relation $m^{d}_{11}= m^{d}_{22}$, we can obtain the celebrated Gatto-Sartori relation for the Cabibbo angle~\cite{Gatto:1968ss}, i.e.
\begin{eqnarray}
\label{eq:gatto}
\frac{m_d}{m_s}\simeq \tan^2\theta_c\,.
\end{eqnarray}

\section{ Global fit of flavor observables}
\label{sec:analysis}

We have already discussed the predictions for the oscillation parameters, summarized in Eq.~(\ref{eq:osc-predictions}).
They are shown in Figs.~\ref{fig:osc-predictions-I} and ~\ref{fig:osc-predictions-II}.
Likewise, the predictions for the absolute neutrino mass scale relevant for neutrinoless double beta decay, tritium beta decays and cosmology were discussed in Fig.~\ref{fig:znbb} and Table~\ref{Tab:mnu_beta_sum}.
Finally, the quark sector prediction for the Cabibbo angle is given in Eq.~(\ref{eq:gatto}).

We now present a global description of all flavor observables in the theory, including the quark and lepton mass parameters as well as the Cabibbo–Kobayashi–Maskawa (CKM) quark mixing parameters.

\subsection{Global Flavour Fit}
\label{sec:global-flavour-fit}

In our numerical analysis, we assume that the fundamental 5-D scale is $k\simeq \Lambda \simeq M_{\text{Pl}}$, with $M_{\text{Pl}}\simeq 2.44\times 10^{18} \text{GeV}$ the reduced Planck mass.
In order to account for the hierarchy between the Planck and the electroweak scales we also set the scale $\Lambda^\prime=ke^{-kL}\simeq 1.5$ TeV.
This allows for the lowest Kaluza-Klein gauge boson resonances (with masses $m_{KK}=3\sim4$ TeV) to be within reach of the LHC experiments.
The Higgs VEV is identified with its \sm value $v\simeq 174$ GeV, and the ratios
$v_{\varphi_{l}}/\Lambda'$, $v_{\sigma_{l}}/\Lambda'$, $v_{\varphi_{\nu}}/\Lambda$, $v_{\rho_{\nu}}/\Lambda$, are all fixed to 0.2, assuming real flavon VEVs.
We now give a set of benchmark values for the bulk mass parameters and coupling constants of the model.
In the lepton sector, we can choose
\begin{eqnarray}
c_l = 0.460\,,~~
c_e = -0.725\,,~~
c_\mu = -0.553\,,~~
c_\tau = -0.117\,,~~
y_e = 1.0\,,~~
y_{\mu} = 1.0\,,~~
y_{\tau} = 1.0\,,~~
\end{eqnarray}
and
\begin{eqnarray}
&&\text{NO}: c_\nu = -0.404\,,~~
y_{\nu1} =
y_{\nu2} =
y_{\nu3} = 1\,,~~
y_{\nu4} = 0.235 +0.0770 i\,,~~
y_{\nu5} = 0.340 +0.0710 i\,,\\
&&\text{IO}: c_\nu = -0.383\,,~~
y_{\nu1} =
y_{\nu2} =
y_{\nu3} = 1\,,~~
y_{\nu4} = -0.354+0.275 i\,,~~
y_{\nu5} = -0.562+0.270 i\,.
\end{eqnarray}
The resulting predictions for neutrino and charged lepton masses as well as lepton mixing parameters are given as part of table~\ref{Tab:fitted_fermion_parameters},
and they reproduce very well current experimental data. The numerical values of the right handed neutrino masses are about $10^{12}\text{GeV}\sim10^{13}$GeV.
For the quark sector we take
\begin{eqnarray}
\nonumber
&c_{UC} = 0.587\,,\quad
c_T = -0.980\,,   \quad
c_u = -0.516\,,  \quad
c_c = -0.555\,, \quad
c_t = 0.966\,,    \quad
c_{ds} = -0.503\,,\quad
c_b = -0.532\,,&\\
&y_u = 6.321\,, \quad
y_t = 6.20\,, \quad
y_u' = 4.00\,,    \quad
y_c' = 1.00\,, \quad
y_t' = 8.30\,,  \quad
y_{ds1} = 4.00\,, \quad
y_{ds2} = 0.892\,, \quad
y_b' = 4.00\,.&
\end{eqnarray}
Thus the numerical fitted results of quark mass matrices are given by
\begin{equation}
m_{u}=\left( \begin{array}{ccc}
0.109 & 0 & 0 \\
0 & 0 & 7.407 \\
29.532 & 0.589 & -145.433-88.418 i\\ \end{array} \right)\,,\quad
m_{d}=\left( \begin{array}{ccc}
0.0198 & 0 & 0 \\
0.0443+0.0769 i & 0.0198 & 0 \\
0 & 0 & 4.180\\ \end{array} \right)\,,
\end{equation}
in GeV units. The fitted values of fermion masses and the mixing parameters are summarized in table~\ref{Tab:fitted_fermion_parameters}.
In particular the fitted CKM matrix is given as
\begin{equation}
  \label{eq:CKM}
V_{\text{CKM}}\simeq\left(
\begin{array}{ccc}
 0.974+0.0175 i ~&-0.0331+0.223 i ~&    -0.00367 \\
0.0329+0.222 i  ~& 0.973-0.0176 i ~&    -0.0359+0.0219 i\\
-0.00010+0.00879 i      ~& 0.0353+0.0215 i      ~&  0.999
\end{array}
\right)\,,
\end{equation}
while the fitted value for the Jarlskog invariant is
\begin{equation}
  \label{eq:Jarlskog}
  J_{CP}^{q}=3.14\times 10^{-5}\,.
\end{equation}
%
\begin{table}[t!]
\centering
\begin{tabular}{|c|c|c|}
\hline
\hline
parameters
& best-fit $\pm$ $1\sigma$ & predictions\\
\hline
$\sin\theta_{12}^{q}$
& 0.22500$\pm0.00100$
& 0.22503  \\
\hline
$\sin\theta_{13}^{q}$
& 0.003675$\pm0.000095$
& 0.003668  \\
\hline
$\sin\theta_{23}^{q}$
& 0.04200$\pm0.00059$
& 0.04205  \\
\hline
$\delta^{q}_{CP}/^{\circ}$
& 66.9$\pm 2$
& 68.2  \\
\hline
$m_{u}~[\text{MeV}]$
& 2.16$^{+0.49}_{-0.26}$
& 2.16   \\
\hline
$m_{c}~[\text{GeV}]$
& 1.27$\pm 0.02$
& 1.27  \\
\hline
$m_{t}~[\text{GeV}]$
& 172.9$\pm 0.4$
& 172.90   \\
\hline
$m_{d}~[\text{MeV}]$
& 4.67$^{+0.48}_{-0.17}$
& 4.21   \\
\hline
$m_{s}~[\text{MeV}]$
& 93$^{+11}_{-5}$
& 93.00   \\
\hline
$m_{b}~[\text{GeV}]$
& 4.18$^{+0.03}_{-0.02}$
& 4.18   \\
\hline
\hline
$\sin^2\theta_{12}^{l} / 10^{-1}$ (NO)
& \multirow{2}{*}{3.20$^{+0.20}_{-0.16}$} & 3.19 \\
$\sin^2\theta_{12}^{l} / 10^{-1}$ (IO)
& & 3.18 \\
\hline
$\sin^2\theta_{23}^{l} / 10^{-1}$ (NO)
& 5.47$^{+0.20}_{-0.30}$ & 5.47  \\
$\sin^2\theta_{23}^{l} / 10^{-1}$ (IO)
& 5.51$^{+0.18}_{-0.30}$ & 5.51 \\
\hline
$\sin^2\theta_{13}^{l}/ 10^{-2}$ (NO)
& 2.160$^{+0.083}_{-0.069}$ & 2.160 \\
$\sin^2\theta_{13}^{l} / 10^{-2}$ (IO)
& 2.220$^{+0.074}_{-0.076}$ & 2.220 \\
\hline
$\delta_{CP}^{l}/\pi$ (NO)
& 1.32$^{+0.21}_{-0.15}$ & 1.567\\
$\delta_{CP}^{l}/\pi$ (IO)
& 1.56$^{+0.13}_{-0.15}$ & 1.571 \\
\hline
$m_{e}~[\text{MeV}]$ & $0.511 \pm 3.1\times 10^{-9}$ & $0.511$ \\
\hline
$m_{\mu}~[\text{MeV}]$ & $105.658 \pm 2.4\times 10^{-6}$ & $105.658$ \\
\hline
$m_{\tau}~[\text{MeV}]$ & $1776.86 \pm 0.12$ & $1776.86$ \\
\hline
$\Delta m_{21}^{2}~[10^{-5}\text{eV}^{2}]$ (NO)
& \multirow{2}{*}{ 7.55$^{+0.20}_{-0.16}$} & \multirow{2}{*}{ 7.55}  \\
$\Delta m_{21}^{2}~[10^{-5}\text{eV}^{2}]$ (IO)
&  & \\
\hline
$|\Delta m_{31}^{2}|~[10^{-3}\text{eV}^{2}]$ (NO)
&  2.50$\pm$0.03 & 2.50 \\
$|\Delta m_{31}^{2}|~[10^{-3}\text{eV}^{2}]$ (IO)
&  2.42$^{+0.03}_{-0.04}$ & 2.42\\
\hline
$\chi^{2}$ (NO) & \multirow{2}{*}{$-$} & 7.65 \\
$\chi^{2}$ (IO) &  & 7.66 \\
\hline
\hline
\end{tabular}
\caption{\label{Tab:fitted_fermion_parameters}
Global warped flavordynamics fit: the neutrino oscillation parameters are taken from the global analysis in~\cite{deSalas:2017kay}, while the quark parameters are obtained from the PDG~\cite{Tanabashi:2018oca}.}
\end{table}
The phenomenological implications of the Randall-Sundrum model with custodial symmetry have been extensively studied in the literature~\cite{Agashe:2003zs,Carena:2007ua,Agashe:2013kxa}.
  The mass scale of the Kaluza-Klein (KK) excitations can be as low as a few TeV consistent with electroweak precision measurements, flavor changing neutral current processes in rare Kaons and B mesons
  as well as lepton flavor violation processes.
  Some simple discrete family symmetries such as $A_4$, $S_4$ have been implemented in a warped extra dimensional setup with custodial symmetry~\cite{Csaki:2008qq,Kadosh:2010rm,Kadosh:2011id}.
  It has been shown that the discrete flavor symmetry can improve over the generic Randall-Sundrum models, and it can greatly weaken bounds from lepton flavor violation, since the lepton doublets are usually assigned to a triplet of the flavor symmetry group and, consequently, their bulk wave functions are universal~\cite{Csaki:2008qq,Kadosh:2010rm,Kadosh:2011id}.
\vskip .1cm
In the present work, we construct a warped model based on the $T'$ flavor symmetry to explain the observed pattern of quark and lepton masses and mixings. Our basic setup is similar to previous flavor models~\cite{Csaki:2008qq,Kadosh:2010rm,Kadosh:2011id}. Therefore we expect that the phenomenology related to KK resonances will be qualitatively similar, the constraints from electroweak precision parameters and measurements of flavor violation in rare decays of Kaons, B mesons and charged leptons can be saturated for KK mass scales of the order of a few TeV.
The flavons $\varphi_l$, $\sigma_l$ and $\varphi_{\nu}$, $\rho_{\nu}$ are localized on the IR and UV branes, respectively.
  Thus the masses of the physical modes of the $\varphi_l$, $\sigma_l$ and $\varphi_{\nu}$, $\rho_{\nu}$ are naturally of order TeV and Planck scale respectively.
  Since all the flavons are gauge singlets, they do not participate in the gauge interactions.
  Their couplings to quarks and leptons arise from higher dimensional operators suppressed by the flavor scale, hence they are typically very small.

\section{High Order Corrections}
\label{sec:highorder}

The above predictions for lepton and quark masses and flavor mixing will receive corrections from higher dimensional operators consistent with the symmetry of the model.
In the following, we analyze the next-to-leading order (NLO) corrections to $\mathcal{L}^{l, \nu, d, u}_{Y}$ in Eqs.~(\ref{eq:potential_charged_lepton},  \ref{eq:potential_neutrino}, \ref{eq:LYd}, \ref{eq:LYu}) and estimate their contributions to the fermion masses and the mixing parameters.

We first discuss the charged lepton sector. The cubic flavon terms in charged lepton interactions are forbidden by the $Z_{4}$ symmetry. Thus the next-to-leading order corrections to the charged lepton masses are coupled with quartic flavon interactions.
Neglecting terms whose contributions can be absorbed into the leading-order Yukawa couplings, we can write down the sub-leading corrections to $\mathcal{L}^{l}_{Y}$ as
\begin{eqnarray}
\nonumber
\delta\mathcal{L}^{l}_{Y}&=&\frac{\sqrt{G}}{\Lambda'^5}\Big[
 x_{e}(\varphi_{l}\varphi_{l}^{\ast}\sigma_l^2\overline{\Psi}_{l})_{\mathbf{1''}}H\Psi_{e}
+x_{\mu}(\varphi_{l}\varphi_{l}^{\ast}\sigma_l^2\overline{\Psi}_{l})_{\mathbf{1'}}H\Psi_{\mu}
+x_{\tau}(\varphi_{l}\varphi_{l}^{\ast}\sigma_l^2\overline{\Psi}_{l})_{\mathbf{1}}H\Psi_{\tau}\\
\label{eq:yuk_l_nlo}&&\quad~~~
+x_{e}'(\varphi_{l}^{\ast 2}\sigma_l^{\ast2}\overline{\Psi}_{l})_{\mathbf{1''}}H\Psi_{e}
+x_{\mu}'(\varphi_{l}^{\ast 2}\sigma_l^{\ast2}\overline{\Psi}_{l})_{\mathbf{1'}}H\Psi_{\mu}
+x_{\tau}'(\varphi_{l}^{\ast 2}\sigma_l^{\ast2}\overline{\Psi}_{l})_{\mathbf{1}}H\Psi_{\tau}
\Big]\delta(y-L)+\text{h.c.}\,.
\end{eqnarray}
The correction to the charged lepton mass matrix is given as
\begin{eqnarray}
\label{eq:ml_nlo}\delta m_{l}=v
\left(\frac{v_{IR}}{\Lambda'}\right)^4\left(\begin{array}{ccc}
0 & \tilde{x}_\mu & \tilde{x}_\tau' \\
\tilde{x}_e' & 0 & \tilde{x}_\tau \\
\tilde{x}_e & \tilde{x}_\mu' & 0
\end{array}\right)\,,
\end{eqnarray}
where $\tilde{x}_{e,\mu,\tau}=\frac{x_{e,\mu,\tau}}{L\Lambda'}f_L(L,c_\ell)f_R(L,c_{e, \mu, \tau})$ and
$\tilde{x}_{e,\mu,\tau}'=\frac{x_{e,\mu,\tau}'}{L\Lambda'}f_L(L,c_\ell)f_R(L,c_{e, \mu, \tau})$.
For simplicity of notation, in Eq.~\eqref{eq:ml_nlo} we have used $v_{IR}$ to represent the VEVs of IR localized flavons $\varphi_l$, $\sigma_l$ and their complex conjugates.
In the neutrino sector, the zero-mode of the right-handed neutrinos is exponentially localized towards the UV brane. As a result, the sub-leading operators involving the IR-localized flavons $\varphi_l$ and $\sigma_l$ are highly suppressed and can be neglected. The NLO corrections to $\mathcal{L}^{\nu}_{Y}$ are of the following form
\begin{eqnarray}
\nonumber\delta\mathcal{L}^{\nu}_{Y}&=&
\frac{\sqrt{G}}{\Lambda'^3}\Big[x_{\nu_{A}}\left((\overline{\Psi}_{l}H\Psi_{\nu})_{\mathbf{3_A}}(\varphi_{l}\varphi_{l}^{\ast})_{\mathbf{3}}\right)_{\mathbf{1}}
+x_{\nu_{S}}\left((\overline{\Psi}_{l}H\Psi_{\nu})_{\mathbf{3_S}}(\varphi_{l}\varphi_{l}^{\ast})_{\mathbf{3}}\right)_{\mathbf{1}}\Big]\delta(y-L)\\
\nonumber&& +
  \frac{1}{2}\frac{\sqrt{G}}{\Lambda^{4}}\Big[x_{\nu_{1}}\big((\overline{N^C}N)_{\mathbf{1''}}f(\varphi_\nu,\rho_\nu)_{\mathbf{1'}}\big)_{\mathbf{1}}+ x_{\nu_{2}}\big((\overline{N^C}N)_{\mathbf{1'}}f(\varphi_\nu,\rho_\nu)_{\mathbf{1''}}\big)_{\mathbf{1}} \\
\label{eq:yuk_nu_nlo}&~&\quad\qquad
 +x_{\nu_{3}}\big((\overline{N^C}N)_{\mathbf{3_{S}}}f(\varphi_\nu,\rho_\nu)_{\mathbf{3_{S}}}\big)_{\mathbf{1}} + x_{\nu_{4}}\big((\overline{N^C}N)_{\mathbf{3_{S}}}f(\varphi_\nu,\rho_\nu)_{\mathbf{3_{A}}}\big)_{\mathbf{1}}\Big]\delta(y)+\text{h.c.}\,,
\end{eqnarray}
where $f(\varphi_\nu,\rho_\nu)$ represent the flavon combinations
$\varphi_\nu^2\rho_\nu\rho_\nu^\ast$,
$\rho_\nu^2\varphi_\nu\varphi_\nu^\ast$,
$\varphi_\nu^\ast\rho_\nu^{\ast3}$ and
$\rho_\nu^\ast\varphi_\nu^{\ast3}$.
As a result, the corrections to the Dirac and Majorana neutrino mass matrices are
\begin{eqnarray}
\delta m_{D}&=&
v\left(\frac{v_{IR}}{\Lambda'}\right)^2\left(\begin{array}{ccc}
2\tilde{x}_{\nu_{S}} & 0 & 0 \\
0 & -\tilde{x}_{\nu_{S}}-\tilde{x}_{\nu_{A}} & 0 \\
0 & 0 & -\tilde{x}_{\nu_{S}}+\tilde{x}_{\nu_{A}}
\end{array}\right)\,,\\
\delta m_{N}&=&\frac{v_{UV}^4}{\Lambda^3}
\left[
 \tilde{x}_{\nu_{1}}\left(\begin{array}{ccc} 0 & 0 & 1 \\ 0 & 1 & 0 \\ 1 & 0 & 0 \\ \end{array} \right)
+\tilde{x}_{\nu_{2}}\left(\begin{array}{ccc} 0 & 1 & 0 \\ 1 & 0 & 0 \\ 0 & 0 & 1 \\ \end{array} \right)
+\frac{\tilde{x}_{\nu_{3}}}{3}
\left(\begin{array}{ccc}
 2 & -1 & -1 \\
 -1 & 2 & -1 \\
 -1 & -1 & 2
\end{array}\right)
+\frac{\tilde{x}_{\nu_{4}}}{6}
\left(\begin{array}{ccc}
 4 & 1 & 1 \\
 1 & -2 & -2 \\
 1 & -2 & -2
\end{array} \right)\right]\,,
\end{eqnarray}
where $\tilde{x}_{\nu_{S,A}}=\frac{x_{\nu_{S,A}}}{L\Lambda'}f_L(L,c_\ell)f_R(L,c_{\nu})$ and $\tilde{x}_{\nu_{i}}=\frac{x_{\nu_{i}}}{L\Lambda}f^2_R(0,c_{\nu})$, and
$v_{UV}$ in $\delta m_N$ denotes the VEVs $v_{\varphi_\nu}$, $v_{\sigma_\nu}$, $v_{\varphi_\nu}^\ast$ and $v_{\sigma_\nu}^\ast$.
To show the contributions of the above subleading terms, we perform a numerical analysis for the case of normal ordered neutrino masses, treating all coupling constants in
  Eq.~\eqref{eq:yuk_l_nlo} and Eq.~\eqref{eq:yuk_nu_nlo} as random complex numbers of absolute value between 0 and 1. We find that the corrections to the lepton masses and mixing parameters are small, and their order of magnitudes are
\begin{eqnarray}
\nonumber
&|\delta m_e| \sim 0.0003{\rm MeV}\,,\quad~~
|\delta m_\mu| \sim 0.04{\rm MeV}\,,\quad~~
|\delta m_\tau| \sim 1{\rm MeV}\,,\quad~~
|\delta(\Delta m^2_{21})| \sim  0.3[10^{-5}{\rm eV}^2]\,,\\
&|\delta(\Delta m^2_{31})| \sim  0.2[10^{-3}{\rm eV}^2]\,,\quad~~
|\delta s_{13}^2|\sim 0.005\,,\quad
|\delta s_{12}^2|\sim 0.06\,,\quad
|\delta s_{23}^2|\sim 0.02\,,\quad
|\delta(\delta_{CP})|\sim 0.03\pi\,.&
\end{eqnarray}
The next-to-leading order corrections to the quark mass terms involve three flavon fields.
Ignoring those terms whose contributions can be absorbed by the leading order operators, the subleading operators of the quark Yukawa interactions are
\begin{eqnarray}
\nonumber \delta\mathcal{L}^u_{Y}&=&
\frac{\sqrt{G}}{\Lambda'^4}\Big[
  x_{c_1}\big((\overline{\Psi}_{UC}{H}\Psi_{c})_{\mathbf{2'}} (\varphi_{l}\sigma_{l}^{2})_{\mathbf{2'}}\big)_{\mathbf{1}}
+ x_{c_2}\big((\overline{\Psi}_{UC}{H}\Psi_{c})_{\mathbf{2'}}(\varphi_{l}^{\ast}\sigma_l^{\ast2})_{\mathbf{2'}}\big)_{\mathbf{1}}
+ x_{t}  \big((\overline{\Psi}_{UC}{H}\Psi_{t})_{\mathbf{2}} (\varphi_{l}\sigma_{l}^{\ast2})_{\mathbf{2''}} \big)_{\mathbf{1}}
\Big]\delta(y-L)\,,\\
\nonumber \delta\mathcal{L}^d_{Y}&=&
\frac{\sqrt{G}}{\Lambda'^4}\Big[
  x_{b_1} \big((\overline{\Psi}_{UC}{H}\Psi_{b})_{\mathbf{2'}} (\varphi_{l}\sigma_{l}^{2})_{\mathbf{2'}}\big)_{\mathbf{1}}
+ x_{b_2} \big((\overline{\Psi}_{UC}{H}\Psi_{b})_{\mathbf{2'}} (\varphi_{l}^{\ast}\sigma_l^{\ast2})_{\mathbf{2'}}\big)_{\mathbf{1}}
+ x_{ds_1}\big((\overline{\Psi}_{T}{H}\Psi_{ds})_{\mathbf{2'}} (\varphi_{l}\sigma_{l}^{2})_{\mathbf{2'}}\big)_{\mathbf{1}}\\
\label{eq:yuk_q_nlo}&&\quad\quad
+ x_{ds_2}\big((\overline{\Psi}_{T}{H}\Psi_{ds})_{\mathbf{2'}} (\varphi_{l}^{\ast}\sigma_l^{\ast2})_{\mathbf{2'}}\big)_{\mathbf{1}}\Big]\delta(y-L)\,.
\end{eqnarray}
Notice that the above operators with the combinations $\varphi_{l}\sigma_{l}^{2}$, $\varphi^{\ast}_{l}\sigma_{l}^{\ast2}$ replaced by $\varphi^3_{l}$ and $\varphi^{\ast3}$ are also allowed by symmetries of the model but their contributions are vanishing for the leading order alignment in Eq.~\eqref{eq:vacuum_lepton}.
The higher terms in Eq.~\eqref{eq:yuk_q_nlo} induce the following corrections to the quark mass matrices,
\begin{eqnarray}
\delta m^{u}=v\left(\frac{v_{IR}}{\Lambda'}\right)^3
\left(\begin{array}{ccc}
  0 & \tilde{x}_{c_1} & \tilde{x}_t \\
  0 & \tilde{x}_{c_2} & 0 \\
  0 & 0 & 0
\end{array}\right) \,,\qquad \delta m^{d}=v\left(\frac{v_{IR}}{\Lambda'}\right)^3
\left(\begin{array}{ccc}
  0 & 0 & \tilde{x}_{b_1} \\
  0 & 0 & \tilde{x}_{b_2} \\
\tilde{x}_{ds_2} & \tilde{x}_{ds_1} & 0 \\
\end{array}\right)
\,,
\end{eqnarray}
with $\tilde{x}_{b_i,c_i,t}=\frac{x_{b_i,c_i,t}}{L\Lambda'}f_L(L,c_{UC})f_R(L,c_{b,c,t})$ and $\tilde{x}_{ds_i}=\frac{x_{ds_i}}{L\Lambda'}f_L(L,c_{T})f_R(L,c_{ds})$.
A numerical analysis similar to the lepton sector is performed, in which all couplings in Eq.~\eqref{eq:yuk_q_nlo} are taken to be random complex numbers with with absolute value in the range of 0 and 1. We find the order of magnitude of the corrections to the quark masses and CKM mixing parameters are
\begin{eqnarray}
\nonumber
&|\delta m_u| \sim 0.4{\rm MeV}\,,\quad
|\delta m_c| \sim 0.2{\rm MeV}\,,\quad
|\delta m_t| \sim 0.002{\rm MeV}\,,\quad
|\delta m_d| \sim 0.04{\rm MeV}\,,\quad
|\delta m_s| \sim 0.1{\rm MeV}\,,\\
&|\delta m_b| \sim 10{\rm MeV}\,,\quad
|\delta(\sin\theta_{13}^q)|\sim 0.0002\,,\quad
|\delta(\sin\theta_{12}^q)|\sim 0.002\,,\quad
|\delta(\sin\theta_{23}^q)|\sim 0.0006\,,\quad
|\delta(\delta_{CP}^q)|\sim 0.009\pi\,.
\end{eqnarray}

\section{ Variant model with bulk Higgs}
\label{sec:bulked_higgs}

In this section, we consider another scheme in which the Higgs field lives in the bulk.
  We assume that the flavons $\varphi_{l}$ and $\sigma_{l}$ are localized on the UV brane, the flavons $\varphi_{\nu}$ and $\rho_{\nu}$ are localized on IR brane, while their vacuum alignments stay the same.
  Since the Higgs field and all fermions live in the bulk in this model, we extend the flavor symmetry to $T'\times Z_3\times Z_8$ in order to forbid unwanted Yukawa terms.
  The transformation properties of fields under $T'\times Z_{3}$ are the same as those in table~\ref{Tab:assignment_lepton} and table~\ref{Tab:assignment_quark}.
  The $Z_{8}$ charge assignments are given as follows
\begin{eqnarray}
\Psi_{l},\Psi_{e},\Psi_{\mu},\Psi_{\tau},\Psi_{\nu}:~\omega_8^{3}\,,\quad
\Psi_{T},\Psi_{c},\Psi_{t},\Psi_{b},\varphi_{l},\sigma_{l}:~\omega_8^{4}\,,\quad
\Psi_{UC},\Psi_{u},\Psi_{ds},H:~1\,,\quad
\varphi_{\nu}:~\omega_8\,,\quad
\rho_{\nu}:~\omega_8^{5}\,,
\end{eqnarray}
where $\omega_8= e^{\pi i/4}$. Within this new setup, the quark and lepton mass terms are still given by Eqs.~(\ref{eq:potential_charged_lepton}, \ref{eq:potential_neutrino}, \ref{eq:LYd}, \ref{eq:LYu}) with $\delta(y)$ and $\delta(y-L)$ interchanged. As a result, both quark and lepton mass matrices are determined to be of the same forms as Eqs.~(\ref{eq:ml}, \ref{eq:mD-mN}, \ref{eq:mass_quark-down}, \ref{eq:mass_quark-up}). Hence the excellent global fit to the observed values of quark and lepton masses and mixing parameters discussed in section~\ref{sec:analysis} can be reproduced for adequately chosen values of the coupling constants and bulk mass parameters. A basic difference is that the right-handed neutrino masses would be as low as 300 GeV in this scenario. The Dirac neutrino Yukawa couplings would be correspondingly smaller so as to keep the correct neutrino masses.

\section{Summary and conclusions}
\label{sec:Conclusions}

We have proposed a realistic five-dimensional warped extension of the standard model where all leptons and quarks propagate in the bulk, see Figs.~\ref{fig:wavef_lepton} and \ref{fig:wavef_quark}.
We have assumed a $T' \otimes Z_{3} \otimes Z_{4}$ family symmetry broken on the branes by flavon fields.
These are also responsible for inducing \lnv as well as \lfv which are therefore connected.
This is a nice feature of the theory, and implies that all the flavor violation needed to account for neutrino oscillations comes entirely from the Majorana seesaw sector.
We have shown that the model provides a consistent scenario for the flavor problem, in which fermion mass hierarchies are accounted for by adequate choices of the bulk mass parameters,
while quark and lepton mixing angles are restricted by the flavor symmetry.
Neutrino masses are generated by the type-I seesaw mechanism, with the Majorana mass terms of the right-handed neutrinos UV--localized, so that the large scale required by the seesaw mechanism
is naturally accommodated in Eq.~(\ref{eq:potential_neutrino}). This is the only fermion mass term that arises from a Yukawa coupling in the UV--brane.
Turning on this Majorana mass term $M_N$ is crucial in order to induce a viable pattern of neutrino masses and mixings.
Without this term neutrinos would be mass-degenerate and unmixed Dirac fermions.
The flavor transformation properties of the heavy Majorana block are dictated by the corresponding flavon fields.
Once these acquire their vevs one gets predictions for neutrino oscillations.
Indeed, neutrino mixing parameters and the Dirac CP violation phase are all described in terms of just two independent parameters.
The resulting predictions for the neutrino oscillation parameters are summarized in Figs.~\ref{fig:osc-predictions-I} and \ref{fig:osc-predictions-II}.

Likewise, our theory predicts a \znbb decay rate within reach of the upcoming generation of experiments, as seen in Fig.~\ref{fig:znbb}.
We have also discussed the predictions for tritium beta decays and cosmology, given in Table~\ref{Tab:mnu_beta_sum}.

Finally, our scheme also provides a good description of the quark sector and the CKM matrix, as seen in Eqs.~(\ref{eq:CKM}, \ref{eq:Jarlskog}), recovering the successful Gatto-Sartori relation for the Cabibbo angle in Eq.~(\ref{eq:gatto}).
In fact, we have actually performed a global flavordynamics fit in our warped scenario, obtaining very good results, presented in table~\ref{Tab:fitted_fermion_parameters}.
We have also studied the higher order corrections, showing that they are small enough to be neglected.
Finally, we have commented on an alternative variant of the model, in which the Higgs field lives in the bulk.
Even though the setup is quite different, the predictions for quark and lepton masses and mixing parameters are kept unchanged.

\acknowledgements
\noindent

This work is supported by the National Natural Science Foundation of China under Grant Nos 11847240, 11975224, 11835013, 11947301 and by the Spanish grants FPA2017-85216-P (AEI/FEDER, UE), PROMETEO/2018/165 (Generalitat Valenciana) and the Spanish Red Consolider MultiDark FPA2017-90566-REDC.

\clearpage

\appendix

\section*{Appendix}

\setcounter{equation}{0}
\renewcommand{\theequation}{\thesection.\arabic{equation}}

\begin{appendix}

\section{\label{sec:Tp_group}Group Theory of $T^{\prime}$}

The $T'$ group is the double covering of the tetrahedral group $A_4$. It has 24 elements which can be generated by two generators $S$ and $T$ obeying the relations\footnote{The $T'$ group can also be equivalently expressed in terms of three generators $S$, $T$ and $R$ with $S^2=R$, $RT=TR$ and $(ST)^{3}=T^3=R^2=1$\cite{Feruglio:2007uu,Ding:2008rj,Liu:2019khw,Lu:2019vgm}.},
\begin{equation}
S^4=(ST)^{3}=T^3=1,~~~S^2T=TS^2\,.
\end{equation}
The $T'$ group has seven inequivalent irreducible representations: three singlets $\mathbf{1}$, $\mathbf{1}'$ and $\mathbf{1}''$, three doublets $\mathbf{2}$, $\mathbf{2}'$ and $\mathbf{2}''$, and one triplet $\mathbf{3}$. The representations $\mathbf{1}'$, $\mathbf{1}''$ and $\mathbf{2}$, $\mathbf{2}''$ are complex conjugated to each other respectively. The two-dimensional representations $\mathbf{2}$, $\mathbf{2}'$ and $\mathbf{2}''$ are faithful representations of $T'$ group, while the odd dimensional representations $\mathbf{1}$, $\mathbf{1}'$, $\mathbf{1}''$ and $\mathbf{3}$ coincide with those of $A_4$.
In the present work we shall adopt the basis of~\cite{Liu:2019khw,Lu:2019vgm}. For the singlet representations, we have
\begin{eqnarray}
\nonumber&&\mathbf{1}: S=T=1\,,\\
\nonumber&&\mathbf{1}': S=1,~~~ T=\omega\,,\\
&&\mathbf{1}'': S=1,~~~ T=\omega^2\,,
\end{eqnarray}
with $\omega=e^{i2\pi/3}$. In the doublet representations, the generators $S$ and $T$ are given by
\begin{eqnarray}
\nonumber&&\mathbf{2}: S=-\frac{1}{\sqrt{3}}\begin{pmatrix}
 i & \sqrt2e^{i\pi/12} \\
 -\sqrt2e^{-i\pi/12} & -i
\end{pmatrix},~~~~~T=\begin{pmatrix}
\omega ~& 0 \\
0 ~& 1
\end{pmatrix}\,,\\
\nonumber&&\mathbf{2}': S=-\frac{1}{\sqrt{3}}\begin{pmatrix}
 i & \sqrt2e^{i\pi/12} \\
 -\sqrt2e^{-i\pi/12} & -i
\end{pmatrix},~~~T=\begin{pmatrix}
\omega^2 & 0 \\
0 & \omega
\end{pmatrix}\,,\\
&&\mathbf{2}'': S=-\frac{1}{\sqrt{3}}\begin{pmatrix}
 i & \sqrt2e^{i\pi/12} \\
 -\sqrt2e^{-i\pi/12} & -i
\end{pmatrix},~~~T=\begin{pmatrix}
1 & 0 \\
0 & \omega^2
\end{pmatrix}\,.
\end{eqnarray}
For the triplet representation $\mathbf{3}$, the generators are
\begin{equation}
S=\frac{1}{3}\begin{pmatrix}
-1 & 2 & 2\\
2 & -1 & 2\\
2 & 2 & -1
\end{pmatrix},~~~~T=\begin{pmatrix}
1 ~& 0 ~& 0\\
0 ~& \omega ~& 0\\
0 ~& 0 ~&\omega^2
\end{pmatrix}\,.
\end{equation}
Notice that due to the choice of complex representation matrices for the real representation $\mathbf{3}$ the conjugate $a^{*}$ of $a\sim\mathbf{3}$ does not transform as $\mathbf{3}$, but rather $(a^{*}_1, a^{*}_3, a^{*}_2)$ transforms as triplet under $T'$. The reason for this
is that $T^{*} = U^T_{3}TU_{3}$ and $S^{*}=U^T_{3}SU_{3}=S$ where $U_{3}$ is the permutation matrix which exchanges the second and third row and column.
Similarly, notice that the irreducible representations $\mathbf{2}$ and $\mathbf{2''}$ are complex conjugated to each other by a unitary transformation
$U_{2}$ with
\begin{equation}
U_{2}=\begin{pmatrix} 0 & -1 \\ 1 & 0 \end{pmatrix},
\end{equation}
i.e, $T_{\mathbf{2}}^{*}=U_{2}^{\dagger}T_{\mathbf{2''}}U_{2}$ and $S_{\mathbf{2}}^{*}=U_{2}^{\dagger}S_{\mathbf{2''}}U_{2}$. Besides, the real doublet representation $\mathbf{2'}$ and its complex conjugation are also related by the unitary transformation $U_{2}$, i.e, $T_{\mathbf{2'}}^{*}=U_{2}^{\dagger}T_{\mathbf{2'}}U_{2}$ and $S_{\mathbf{2'}}^{*}=U_{2}^{\dagger}S_{\mathbf{2'}}U_{2}$. Thus we have
\begin{eqnarray}
\nonumber
&&b=(b_{1},b_{2})^{T}\sim  \mathbf{2},\qquad \rightarrow \qquad (-b_{2}^{*},b_{1}^{*})^{T}\sim \mathbf{2''}\,\\
\nonumber
&&b=(b_{1},b_{2})^{T}\sim  \mathbf{2''},\qquad \rightarrow \qquad (-b_{2}^{*},b_{1}^{*})^{T}\sim \mathbf{2}\,\\
&&b=(b_{1},b_{2})^{T}\sim  \mathbf{2'},\qquad \rightarrow \qquad (-b_{2}^{*},b_{1}^{*})^{T}\sim \mathbf{2'}\,.
\end{eqnarray}
In the following, we collect the Clebsch-Gordan coefficients for the decomposition of product representations in our basis, all the results are taken from~\cite{Liu:2019khw,Lu:2019vgm}. We use $\alpha_i$ to indicate the elements of the first representation of the product, $\beta_i $ to indicate those of the second representation. For convenience, we shall denote $\mathbf{1}\equiv\mathbf{1}^0$, $\mathbf{1}'\equiv\mathbf{1}^{1}$, $\mathbf{1}''\equiv\mathbf{1}^{2}$ for singlet representations and $\mathbf{2}\equiv\mathbf{2}^0$, $\mathbf{2}'\equiv\mathbf{2}^{1}$, $2''\equiv\mathbf{2}^{2}$ for the doublet representations.

The contraction rules involving singlets representations in the product are as follows,
\begin{eqnarray}
\mathbf{1}^a\otimes\mathbf{1}^b &=& \mathbf{1}^{a+b~(\text{mod}~3)} \sim \alpha\beta \,, \\
\mathbf{1}^a\otimes\mathbf{2}^b &=& \mathbf{2}^{a+b~(\text{mod}~3)} \sim \left(\begin{array}{c} \alpha\beta_1\\
\alpha\beta_2 \\ \end{array}\right) \,,\\
\mathbf{1'}\otimes\mathbf{3} &=& \mathbf{3} \sim \left(\begin{array}{c} \alpha\beta_3\\
\alpha\beta_1 \\
\alpha\beta_2 \end{array}\right) \,, \\
\mathbf{1''}\otimes\mathbf{3} &=& \mathbf{3} \sim \left(\begin{array}{c} \alpha\beta_2\\
\alpha\beta_3 \\
\alpha\beta_1 \end{array}\right) \,,
\end{eqnarray}
where $a, b=0, 1, 2$. The contraction rules for the products of two doublet representations are
\begin{eqnarray}
  \mathbf{2}\otimes\mathbf{2}=\mathbf{2'}\otimes \mathbf{2''}=\mathbf{3}\oplus\mathbf{1'} ~&~~\text{with}~~&~\left\{
\begin{array}{l}
\mathbf{1'}\sim \alpha_1\beta_2-\alpha_2\beta_1 \\ [0.1in]
\mathbf{3}\sim
\left(\begin{array}{c} e^{i \pi /6}\alpha_2\beta_2 \\
\frac{1}{\sqrt{2}}e^{i 7\pi /12}(\alpha_1\beta_2+\alpha_2\beta_1)  \\
\alpha_1\beta_1 \end{array}\right)
\end{array}
\right. \\
\mathbf{2}\otimes\mathbf{2'}=\mathbf{2''}\otimes\mathbf{2''}=\mathbf{3}\oplus\mathbf{1''}~&~~\text{with}~~&~\left\{
\begin{array}{l}
\mathbf{1''}\sim \alpha_1\beta_2-\alpha_2\beta_1\\ [0.1in]
\mathbf{3}\sim
 \left(\begin{array}{c} \alpha_1\beta_1\\
 e^{i \pi /6}\alpha_2\beta_2 \\
 \frac{1}{\sqrt{2}}e^{i 7\pi /12}(\alpha_1\beta_2+\alpha_2\beta_1) \end{array}\right)
\end{array}
\right. \\
\mathbf{2}\otimes\mathbf{2''}=\mathbf{2'}\otimes\mathbf{2'}=\mathbf{3}\oplus\mathbf{1} ~&~~\text{with}~~&~\left\{
\begin{array}{l}
\mathbf{1} \sim \alpha_1\beta_2-\alpha_2\beta_1 \\ [0.1in]
\mathbf{3}\sim
\left(\begin{array}{c}  \frac{1}{\sqrt{2}}e^{i 7\pi /12}(\alpha_1\beta_2+\alpha_2\beta_1)\\
 \alpha_1\beta_1\\
 e^{i \pi /6}\alpha_2\beta_2  \end{array}\right)
\end{array}
\right.
\end{eqnarray}
The products of doublet and triplet representations are decomposed as follows,
\begin{eqnarray}
\mathbf{2}\otimes\mathbf{3}=\mathbf{2}\oplus\mathbf{2'}\oplus\mathbf{2''} ~&~~\text{with}~~&~\left\{
\begin{array}{l}
\mathbf{2}\sim
 \left(\begin{array}{c} \alpha_1\beta_1-\sqrt{2}e^{i 7\pi /12}\alpha_2\beta_2 \\
 -\alpha_2\beta_1+\sqrt{2}e^{i 5\pi / 12}\alpha_1\beta_3 \end{array}\right)  \\ [0.1in]
\mathbf{2'}\sim
\left(\begin{array}{c} \alpha_1\beta_2-\sqrt{2}e^{i 7\pi /12}\alpha_2\beta_3 \\
-\alpha_2\beta_2+\sqrt{2}e^{i 5\pi / 12}\alpha_1\beta_1  \end{array}\right) \\ [0.1in]
\mathbf{2''}\sim
\left(\begin{array}{c} \alpha_1\beta_3-\sqrt{2}e^{i 7\pi /12}\alpha_2\beta_1 \\
-\alpha_2\beta_3+\sqrt{2}e^{i 5\pi / 12}\alpha_1\beta_2  \end{array}\right) \\ [0.1in]
\end{array}
\right. \\
\mathbf{2'}\otimes\mathbf{3}=\mathbf{2}\oplus\mathbf{2'}\oplus\mathbf{2''} ~&~~\text{with}~~&~\left\{
\begin{array}{l}
\mathbf{2}\sim
\left(\begin{array}{c} \alpha_1\beta_3-\sqrt{2}e^{i 7\pi /12}\alpha_2\beta_1 \\
-\alpha_2\beta_3+\sqrt{2}e^{i 5\pi / 12}\alpha_1\beta_2  \end{array}\right)  \\ [0.1in]
\mathbf{2'}\sim
 \left(\begin{array}{c} \alpha_1\beta_1-\sqrt{2}e^{i 7\pi /12}\alpha_2\beta_2 \\
 -\alpha_2\beta_1+\sqrt{2}e^{i 5\pi / 12}\alpha_1\beta_3 \end{array}\right) \\ [0.1in]
\mathbf{2''}\sim
\left(\begin{array}{c} \alpha_1\beta_2-\sqrt{2}e^{i 7\pi /12}\alpha_2\beta_3 \\
-\alpha_2\beta_2+\sqrt{2}e^{i 5\pi / 12}\alpha_1\beta_1  \end{array}\right) \\ [0.1in]
\end{array}
\right. \\
\mathbf{2''}\otimes\mathbf{3}=\mathbf{2}\oplus\mathbf{2'}\oplus\mathbf{2''} ~&~~\text{with}~~&~\left\{
\begin{array}{l}
\mathbf{2}\sim
\left(\begin{array}{c} \alpha_1\beta_2-\sqrt{2}e^{i 7\pi /12}\alpha_2\beta_3 \\
-\alpha_2\beta_2+\sqrt{2}e^{i 5\pi / 12}\alpha_1\beta_1  \end{array}\right)  \\ [0.1in]
\mathbf{2'}\sim
\left(\begin{array}{c} \alpha_1\beta_3-\sqrt{2}e^{i 7\pi /12}\alpha_2\beta_1 \\
-\alpha_2\beta_3+\sqrt{2}e^{i 5\pi / 12}\alpha_1\beta_2  \end{array}\right) \\ [0.1in]
\mathbf{2''}\sim
 \left(\begin{array}{c} \alpha_1\beta_1-\sqrt{2}e^{i 7\pi /12}\alpha_2\beta_2 \\
 -\alpha_2\beta_1+\sqrt{2}e^{i 5\pi / 12}\alpha_1\beta_3 \end{array}\right) \\ [0.1in]
\end{array}
\right. \\
\end{eqnarray}
Finally the contractions of two triplets are given by
\begin{eqnarray}
\mathbf{3}\otimes\mathbf{3}=\mathbf{3}_S\oplus\mathbf{3}_A\oplus\mathbf{1}\oplus\mathbf{1'}\oplus\mathbf{1''} ~&~~\text{with}~~&~\left\{
\begin{array}{l}
\mathbf{3}_S\sim
 \left(\begin{array}{c} 2\alpha_1\beta_1 - \alpha_2\beta_3 - \alpha_3\beta_2 \\
 2\alpha_3\beta_3 - \alpha_1\beta_2 - \alpha_2\beta_1  \\
 2\alpha_2\beta_2 - \alpha_1\beta_3 - \alpha_3\beta_1 \end{array}\right) \\ [0.1in]
\mathbf{3}_A\sim
 \left(\begin{array}{c} \alpha_2\beta_3 - \alpha_3\beta_2 \\
\alpha_1\beta_2 - \alpha_2\beta_1  \\
 \alpha_3\beta_1 - \alpha_1\beta_3 \end{array}\right) \\ [0.1in]
\mathbf{1} \sim \alpha_1\beta_1 + \alpha_2\beta_3 + \alpha_3\beta_2 \\ [0.1in]
\mathbf{1'} \sim \alpha_3\beta_3 + \alpha_1\beta_2 + \alpha_2\beta_1 \\ [0.1in]
\mathbf{1''} \sim \alpha_2\beta_2 + \alpha_1\beta_3 + \alpha_3\beta_1 \\ [0.1in]
\end{array}
\right.
\end{eqnarray}

\section{\label{sec:XD}5-D profiles of Higgs and fermion fields}

We formulate our model in the framework of Randall-Sundrum model~\cite{Randall:1999ee}, assuming the bulk of our model to be a slice of $\mathrm{AdS}_5$ with curvature radius $1/k$.
The extra dimension $y$ is compactified, and the two 3-branes with opposite tension are located at $y = 0$, the UV brane, and $y=L$, the IR brane. The bulk metric is non-factorizable,
\begin{equation}
ds^2=e^{-2ky}\eta_{\mu\nu}dx^{\mu}dx^{\nu}-dy^2\,.
\end{equation}
In this paper we adopt the zero mode approximation which identifies the standard model fields with zero modes of corresponding 5-D fields.
If the Higgs field lives in the bulk, its Kaluza-Klein decomposition is~\cite{Cacciapaglia:2006mz}
\begin{equation}
H(x^\mu,y)= H(x^\mu) \frac{f_H(y)}{\sqrt{L}} + \text{heavy KK Modes}\,,
\end{equation}
where $f_H(y)$ is the zero mode profile,
%
\begin{equation}
f_H(y)=\sqrt{\frac{2 k L (1-\beta )}{1-e^{-2(1-\beta )k L}}}e^{kL}e^{(2-\beta)k(y-L)}\,,
\end{equation}
with $\beta=\sqrt{4+m^2_H/k^2}$ and $m_H$ is the bulk mass of the Higgs field.
For 5-D fermion fields, the three families of leptons and quarks and their $SU(2)_L\otimes SU(2)_{R}$ assignments are given as follows,
\begin{eqnarray}
\label{eq:leptons-5D}\Psi_{\ell_i}&=&\left(\begin{array}{c}
\nu_{ i}^{[++]}\\
e_{i}^{[++]}
\end{array}\right)\sim (\mathbf{2},\mathbf{1})\,,\qquad
\Psi_{e_i}=\left(\begin{array}{c}
\tilde{\nu}_{i}^{\,[+-]}\\
e_{ i}^{[--]}
\end{array}\right)\sim(\mathbf{1},\mathbf{2})\,,\qquad
\Psi_{\nu_i}=\left(\begin{array}{c}
N_{i}^{[--]}\\
\tilde{e}_{i}^{\,[+-]}
\end{array}\right)\sim(\mathbf{1},\mathbf{2})\,,\\
\label{eq:quarks-5D}\Psi_{Q_i}&=&\left(\begin{array}{c}
u_{i}^{[++]}\\
d_{i}^{[++]}
\end{array}\right)\sim (\mathbf{2},\mathbf{1})\,,\qquad
\Psi_{d_i}=\left(\begin{array}{c}
\tilde{u}_{i}^{\,[+-]}\\
d_{i}^{[--]}
\end{array}\right)\sim(\mathbf{1},\mathbf{2})\,,\qquad
\Psi_{u_i}=\left(\begin{array}{c}
u_{i}^{[--]}\\
\tilde{d}_{i}^{\,[+-]}
\end{array}\right)\sim(\mathbf{1},\mathbf{2})\,.
\end{eqnarray}
where the two signs in the bracket indicate Neumann ($+$) or Dirichlet ($-$) boundary conditions (BCs) for the left-handed component of the corresponding field on UV and IR branes respectively.
The Kaluza-Klein decomposition for the two different BCs are
\begin{eqnarray}
&&\psi^{[++]}(x^\mu,y)=\frac{e^{2ky}}{\sqrt{L}}\Big\{\psi_L(x^\mu)f_L(y,c_{L})+\text{heavy KK modes}\Big\}\,,\\
&&\psi^{[--]}(x^\mu,y)=\frac{e^{2ky}}{\sqrt{L}}\Big\{\psi_R(x^\mu)f_R(y,c_{R})+\text{heavy KK modes}\Big\}\,,
\end{eqnarray}
with $\psi$=$\nu_i,e_i, N_i, u_i,d_i$.
The 5-D fields with $[++]$ BCs only have left-handed zero modes, and those with {$[--]$} BCs only have right-handed zero modes. The functions
$f_L(y,c_{L})$ and $f_R(y,c_{R})$ are the zero mode profiles~\cite{Gherghetta:2000qt,Grossman:1999ra,Huber:2001ug}
\begin{equation}
f_L(y,c_{L})=\sqrt{\frac{(1-2 c_{L})kL}{e^{(1-2c_{L})kL}-1}} e^{-c_{L}ky}\,,
\hspace{1cm}
f_R(y,c_R)=\sqrt{\frac{(1+2c_R)kL}{e^{(1+2c_R)kL}-1}} e^{ c_R ky}\,,
\end{equation}
where $c_{L}$ and $c_{R}$ represent the bulk mass of the 5-D fermions in units of $k$.

\section{\label{sec:vacuum_alignment}Vacuum Alignment}

In this section, we will investigate the vacuum alignment of the flavon fields $\varphi_{l}$, $\sigma_{l}$, $\varphi_{\nu}$ and $\rho_{\nu}$.
At the IR brane $y=L$, the scalar potential invariant under the flavor symmetry $T'$ and the auxiliary symmetry $Z_3\times Z_4$ takes the following form
\begin{eqnarray}
\nonumber V_{IR}&=&
M_{\varphi}^{2}(\varphi_{l}\varphi_{l}^{*})_{\mathbf{1}}
+M_{\sigma}^{2}\sigma_{l}\sigma_{l}^{*}
+f_{1}(\varphi_{l}\varphi_{l})_{\mathbf{1}'}(\varphi_{l}^{*}\varphi_{l}^{*})_{\mathbf{1}''}
+f_{2}e^{-i\pi/6}((\varphi_{l}\varphi_{l})_{\mathbf{3}}(\varphi_{l}^{*}\varphi_{l}^{*})_{\mathbf{3}})_{\mathbf{1}}
+f_{3}\sigma^2_{l}\sigma_{l}^{*2}
+f_{4}(\varphi_{l}\varphi_{l}^\ast)_{\mathbf{1}}\sigma_{l}\sigma_{l}^\ast
\\
&=&
M_{\varphi}^{2}(\varphi_{l1}\varphi_{l1}^\ast+\varphi_{l2}\varphi_{l2}^\ast)+
M_{\sigma}^{2}\sigma_{l}\sigma_{l}^\ast+
f_2(\varphi_{l1}\varphi_{l1}^\ast+\varphi_{l2}\varphi_{l2}^\ast)^2+
f_3\sigma_{l}^2\sigma_{l}^{\ast2}
+f_4\sigma_{l}\sigma_{l}^\ast(\varphi_{l1}\varphi_{l1}^\ast+\varphi_{l2}\varphi_{l2}^\ast)\,,
\end{eqnarray}
where the parameters $M_{\varphi}^{2}$, $M_{\sigma}^{2}$, $f_{1}$, $f_{2}$, $f_{3}$ and $f_{4}$ are real free parameters. For the field configuration
\begin{equation}
\braket{\varphi_{l}}=(1,0)v_{\varphi_{l}}\,,\qquad\braket{\sigma_{l}}=v_{\sigma_{l}}\,,
\end{equation}
the minimum conditions of the IR potential read
\begin{equation}
\begin{aligned}
\frac{\partial V_{UV}}{\partial \varphi_{l1}^{*}}&=v_{\varphi_l}\big( M_\varphi^2+2f_2v_{\varphi_l}v_{\varphi_l}^\ast+f_4v_{\sigma_l}v_{\sigma_l}^\ast \big)=0\,,\\
\frac{\partial V_{UV}}{\partial \varphi_{l2}^{*}}&=0\,,\\
\frac{\partial V_{UV}}{\partial \sigma_{l}^{*}}&=v_{\sigma_l}\big( M_\sigma^2+2f_3v_{\sigma_l}v_{\sigma_l}^\ast+f_4v_{\varphi_l}v_{\varphi_l}^\ast  \big)=0\,.
\end{aligned}
\end{equation}
and the solution is
\begin{equation}
|v_{\varphi_{l}}|^2=\frac{f_4M_{\sigma}^{2} - 2 f_3M_{\varphi}^{2}}{4f_{2}f_3-f_{4}^2}\,,\qquad
|v_{\sigma_{l}}|^2= \frac{f_4M_{\varphi}^{2}-2f_2M_{\sigma}^{2}}{4f_{2}f_3-f_{4}^2}\,.
\end{equation}

At the UV brane $y=0$, the most general renormalizable scalar potential $V_{UV}$ for the flavon fields $\varphi_{\nu}$ and $\rho_{\nu}$ is given as
\begin{eqnarray}
\nonumber
V_{UV}&=&M_{1}(\varphi_{\nu}\varphi_{\nu}^{*})_{\mathbf{1}}+M_{2}(\rho_{\nu}\rho_{\nu}^{*})_{\mathbf{1}}
+g_{1}\left((\varphi_{\nu}\varphi_{\nu})_{\mathbf{1}}(\varphi_{\nu}^{*}\varphi_{\nu}^{*})_{\mathbf{1}}\right)_{\mathbf{1}}
+g_{2}\left((\varphi_{\nu}\varphi_{\nu})_{\mathbf{1'}}(\varphi_{\nu}^{*}\varphi_{\nu}^{*})_{\mathbf{1''}}\right)_{\mathbf{1}}
+g_{3}\left((\varphi_{\nu}\varphi_{\nu})_{\mathbf{1''}}(\varphi_{\nu}^{*}\varphi_{\nu}^{*})_{\mathbf{1'}}\right)_{\mathbf{1}}  \\
&~&
\nonumber
+g_{4}\left((\varphi_{\nu}\varphi_{\nu})_{\mathbf{3_{S}}}(\varphi_{\nu}^{*}\varphi_{\nu}^{*})_{\mathbf{3_{S}}}\right)_{\mathbf{1}}
+g_{5}\left((\varphi_{\nu}\varphi_{\nu}^{*})_{\mathbf{1}}(\rho_{\nu}\rho_{\nu}^{*})_{\mathbf{1}}\right)_{\mathbf{1}}
+g_{6}\left((\varphi_{\nu}\varphi_{\nu}^{*})_{\mathbf{1'}}(\rho_{\nu}\rho_{\nu}^{*})_{\mathbf{1''}}\right)_{\mathbf{1}}
+g_{7}\left((\varphi_{\nu}\varphi_{\nu}^{*})_{\mathbf{1''}}(\rho_{\nu}\rho_{\nu}^{*})_{\mathbf{1'}}\right)_{\mathbf{1}}
  \\
&~&
\nonumber
+g_{8}\left((\varphi_{\nu}\varphi_{\nu}^{*})_{\mathbf{3_{S}}}(\rho_{\nu}\rho_{\nu}^{*})_{\mathbf{3_{S}}}\right)_{\mathbf{1}}
+g_{9}\left((\varphi_{\nu}\varphi_{\nu}^{*})_{\mathbf{3_{S}}}(\rho_{\nu}\rho_{\nu}^{*})_{\mathbf{3_{A}}}\right)_{\mathbf{1}}
+g_{10}\left((\varphi_{\nu}\varphi_{\nu}^{*})_{\mathbf{3_{A}}}(\rho_{\nu}\rho_{\nu}^{*})_{\mathbf{3_{S}}}\right)_{\mathbf{1}}
+g_{11}\left((\varphi_{\nu}\varphi_{\nu}^{*})_{\mathbf{3_{A}}}(\rho_{\nu}\rho_{\nu}^{*})_{\mathbf{3_{A}}}\right)_{\mathbf{1}} \\
&~&
\nonumber
+g_{12}\left((\rho_{\nu}\rho_{\nu})_{\mathbf{1}}(\rho_{\nu}^{*}\rho_{\nu}^{*})_{\mathbf{1}}\right)_{\mathbf{1}}
+g_{13}\left((\rho_{\nu}\rho_{\nu})_{\mathbf{1'}}(\rho_{\nu}^{*}\rho_{\nu}^{*})_{\mathbf{1''}}\right)_{\mathbf{1}}
+g_{14}\left((\rho_{\nu}\rho_{\nu})_{\mathbf{1''}}(\rho_{\nu}^{*}\rho_{\nu}^{*})_{\mathbf{1'}}\right)_{\mathbf{1}}
+g_{15}\left((\rho_{\nu}\rho_{\nu})_{\mathbf{3_{S}}}(\rho_{\nu}^{*}\rho_{\nu}^{*})_{\mathbf{3_{S}}}\right)_{\mathbf{1}}\\
      &~&
\nonumber
+\Big[g_{16}\left((\varphi_{\nu}\varphi_{\nu})_{\mathbf{1}}(\rho_{\nu}^{*}\rho_{\nu}^{*})_{\mathbf{1}}\right)_{\mathbf{1}}
+g_{17}\left((\varphi_{\nu}\varphi_{\nu})_{\mathbf{1'}}(\rho_{\nu}^{*}\rho_{\nu}^{*})_{\mathbf{1''}}\right)_{\mathbf{1}}
+g_{18}\left((\varphi_{\nu}\varphi_{\nu})_{\mathbf{1''}}(\rho_{\nu}^{*}\rho_{\nu}^{*})_{\mathbf{1'}}\right)_{\mathbf{1}}\\
&~&
~~+g_{19}\left((\varphi_{\nu}\varphi_{\nu})_{\mathbf{3_{S}}}(\rho_{\nu}^{*}\rho_{\nu}^{*})_{\mathbf{3_{S}}}\right)_{\mathbf{1}}+\text{h.c}\Big]
\,,
\end{eqnarray}
where the coupling parameters $M_{1,2}$ and $g_{1 \rightarrow 15}$ are real, while the remaining coupling parameters $g_{16,17,18,19}$ are complex. For the desired flavon vacuum alignments
\begin{equation}
\braket{\varphi_{\nu}}=(1,-2\omega^{2},-2\omega)v_{\varphi_{\nu}}\,,\quad \braket{\rho_{\nu}}=(1,-2\omega,-2\omega^{2})v_{\rho_{\nu}}\,,
\end{equation}
we find the minimization conditions are
\begin{eqnarray}
\nonumber
\frac{\partial V_{UV}}{\partial \varphi_{\nu 1}^{*}}&=&A_{1}-12(3g_{8}-2i\sqrt{3}g_{10})v_{\varphi_{\nu}}|v_{\rho_{\nu}}|^{2}-72g_{19}^{*}v_{\varphi_{\nu}}^{*}v_{\rho_{\nu}}^{2}=0\,,\\
\nonumber
\frac{\partial V_{UV}}{\partial \varphi_{\nu 2}^{*}}&=&-2\omega^{2}\left[A_{1}+3(6\omega  g_{8}-(3+i\sqrt{3})g_{10})v_{\varphi_{\nu}}|v_{\rho_{\nu}}|^{2}+36\omega g_{19}^{*}v_{\varphi_{\nu}}^{*}v_{\rho_{\nu}}^{2}\right]=0\,,\\
\nonumber
\frac{\partial V_{UV}}{\partial \varphi_{\nu 3}^{*}}&=&-2\omega\left[A_{1}+3(6\omega^{2}  g_{8}+(3-i\sqrt{3})g_{10})v_{\varphi_{\nu}}|v_{\rho_{\nu}}|^{2}+36\omega^{2} g_{19}^{*}v_{\varphi_{\nu}}^{*}v_{\rho_{\nu}}^{2}\right]=0\,,\\
\nonumber
\frac{\partial V_{UV}}{\partial \rho_{\nu 1}^{*}}&=&A_{2}-12(3g_{8}+2i\sqrt{3}g_{9})v_{\rho_{\nu}}|v_{\varphi_{\nu}}|^{2}-72g_{19}v_{\rho_{\nu}}^{*}v_{\varphi_{\nu}}^{2}=0\,,\\
\nonumber
\frac{\partial V_{UV}}{\partial \rho_{\nu 2}^{*}}&=&-2\omega\left[ A_{2}+3(6\omega^{2}g_{8}-(3-i\sqrt{3})g_{9})v_{\rho_{\nu}}|v_{\varphi_{\nu}}|^{2}+36\omega^{2}g_{19}v_{\rho_{\nu}}^{*}v_{\varphi_{\nu}}^{2}\right]=0\,,\\
\frac{\partial V_{UV}}{\partial \rho_{\nu 3}^{*}}&=&-2\omega^{2}\left[ A_{2}+3(6\omega g_{8}+(3+i\sqrt{3})g_{9})v_{\rho_{\nu}}|v_{\varphi_{\nu}}|^{2}+36\omega g_{19}v_{\rho_{\nu}}^{*}v_{\varphi_{\nu}}^{2}\right]=0\,.
\end{eqnarray}
with
\begin{eqnarray}
\nonumber  A_{1}&=&M_{1}v_{\varphi_{\nu}}+18(g_{1}+4g_{4})v_{\varphi_{\nu}}|v_{\varphi_{\nu}}|^{2}+9g_{5}v_{\varphi_{\nu}}|v_{\rho_{\nu}}|^{2}+18 g_{16}^{*}v_{\varphi_{\nu}}^{*}v_{\rho_{\nu}}^{2}\,,\\
  A_{2}&=&M_{2}v_{\rho_{\nu}}+18g_{16} v_{\rho_{\nu}}^{*}v_{\varphi_{\nu}}^{2}+9g_{5}v_{\rho_{\nu}}|v_{\varphi_{\nu}}|^{2}+18(g_{12}+4g_{15})v_{\rho_{\nu}}|v_{\rho_{\nu}}|^{2}\,.
\end{eqnarray}
From above equations, we find that the non-trivial solutions can be achieved if the couplings $g_{8,9,10,19}$ are vanishing.
Under such assumptions, the minimization conditions are simplified into
\begin{eqnarray}
\nonumber
&&\frac{\partial V_{UV}}{\partial \varphi_{\nu 1}^{*}}=A_{1}=0\,,\quad
\frac{\partial V_{UV}}{\partial \varphi_{\nu 2}^{*}}=-2\omega^{2}A_{1}=0\,,\quad
\frac{\partial V_{UV}}{\partial \varphi_{\nu 3}^{*}}=-2\omega A_{1}=0\,,\\
&&\frac{\partial V_{UV}}{\partial \rho_{\nu 1}^{*}}=A_{2}=0\,,\quad
\frac{\partial V_{UV}}{\partial \rho_{\nu 2}^{*}}=-2\omega A_{2}\,,\quad
\frac{\partial V_{UV}}{\partial \rho_{\nu 3}^{*}}=-2\omega^{2} A_{2}\,.
\end{eqnarray}
One sees that the assumed vacuum alignment of the flavon fields can, indeed, be achieved within certain regions of parameter space.

\end{appendix}

\providecommand{\href}[2]{#2}\begingroup\raggedright\endgroup

\end{document}